\documentclass[letterpaper, 11pt]{article}
\pdfoutput=1

\usepackage{authblk}

\usepackage[table]{xcolor}

\usepackage{breakcites}

\usepackage{natbib}
\defcitealias{Mathematica}{Mathematica}

\usepackage{
  amsmath, amsthm, amssymb, mathtools, dsfont, units,   % Math typesetting
  graphicx, wrapfig, subfig, float,                     % Figures and graphics formatting
  listings, color, inconsolata, pythonhighlight,        % Code formatting
  fancyhdr, sectsty, hyperref, enumerate, enumitem }   	% Headers/footers, section fonts, links, lists

\usepackage[ruled,vlined]{algorithm2e}

\usepackage[left=1.2in, right=1.2in, top=1in, bottom=0.85in, headsep=.2in]{geometry}

\usepackage[bottom]{footmisc}

\renewcommand{\baselinestretch}{1.08}

\allowdisplaybreaks

\sectionfont{\fontsize{14}{15}\selectfont}
\subsectionfont{\normalsize}

\hypersetup{colorlinks=true, linkcolor=blue, citecolor=blue}

\renewenvironment{abstract}
{\small\begin{quote}\noindent \par{\bf \abstractname.}}
{\noindent\end{quote}}

\usepackage{caption}
\definecolor{redcell}{HTML}{ffd4d4}
\definecolor{bluecell}{HTML}{dee7ff}
\definecolor{graycell}{HTML}{dedede}

\newcommand\numberthis{\addtocounter{equation}{1}\tag{\theequation}}

\let\bg\boldsymbol

\let\mc\mathcal

\newcommand{\R}{\mathbb{R}}   % Real numbers
\newcommand{\E}{\mathbb{E}}     % Expectation, e.g. $\E(X)$
\newcommand{\var}{\text{Var}}   % Variance, e.g. $\var(X)$
\newcommand{\tth}{\text{th}}	% Non-italicized 'th', e.g. $n^\tth$
\newcommand{\1}{\mathds{1}}		% Indicator function, e.g. $\1_A$

\newcommand{\LL}{\text{LL}}

\newcommand{\own}{\text{own}}
\newcommand{\other}{\text{other}}

\DeclareMathOperator*{\argmin}{argmin}    % Argmin, e.g. $\argmin_{x\in[0,1]} f(x)$
\DeclareMathOperator*{\dv}{d\!}           % Non-italicized 'with respect to', e.g. $\int f(x) \dv x$
\theoremstyle{definition}
\newtheorem{theorem}{Theorem}[section]

\graphicspath{{Figures/}{../Figures/}}

\begin{document}

% --- Title and authors ---

\title{\large\bfseries\MakeUppercase{An empirical Bayes approach to estimating dynamic}\\ \MakeUppercase{models of co-regulated gene expression}}
\uppercase{\author[1]{Sara Venkatraman}
\author[1]{Sumanta Basu}
\author[2]{Andrew G. Clark}
\author[1,2]{\\Sofie Delbare}
\author[3]{Myung Hee Lee}
\author[1]{Martin T. Wells}}
\affil[1]{Department of Statistics and Data Science, Cornell University}
\affil[2]{Department of Molecular Biology and Genetics, Cornell University}
\affil[3]{Center for Global Health, Department of Medicine, Weill Cornell Medical College}
\date{\vspace{-5ex}}
\maketitle

% --- Main content ---

% TeX root = ../main.tex

\begin{abstract}
Time-course gene expression datasets provide insight into the dynamics of complex biological processes, such as immune response and organ development. It is of interest to identify genes with similar temporal expression patterns because such genes are often biologically related. However, this task is challenging due to the high dimensionality of these datasets and the nonlinearity of gene expression time dynamics. We propose an empirical Bayes approach to estimating ordinary differential equation (ODE) models of gene expression, from which we derive a similarity metric between genes called the Bayesian lead-lag $R^2$ ($\LL R^2$). Importantly, the calculation of the $\LL R^2$ leverages biological databases that document known interactions amongst genes; this information is automatically used to define informative prior distributions on the ODE model's parameters. As a result, the $\LL R^2$ is a biologically-informed metric that can be used to identify clusters or networks of functionally-related genes with co-moving or time-delayed expression patterns. We then derive data-driven shrinkage parameters from Stein's unbiased risk estimate that optimally balance the ODE model's fit to both data and external biological information. Using real gene expression data, we demonstrate that our methodology allows us to recover interpretable gene clusters and sparse networks. These results reveal new insights about the dynamics of biological systems.
\end{abstract}

% TeX root = ../main.tex

\section{Introduction}\label{sec:intro}

Time-course gene expression datasets are an essential resource for querying the dynamics of complex biological processes, such as immune response, disease progression, and organ development \citep{yosef2011impulse, bar2012studying, purvis2013encoding}. Such datasets, now abundantly available through techniques such as whole-genome RNA sequencing, consist of gene expression measurements for thousands of an organism's genes at a few (typically 5-20) time points. Experimental evidence has revealed that groups of genes exhibiting similar temporal expression patterns are often biologically associated \citep{eisen1998cluster}. For instance, such genes may be co-regulated by the same \textit{transcription factors} \citep{tavazoie1999systematic}: proteins that directly control gene expression, which ultimately contributes to changes in cellular function. Identifying clusters or networks of genes with related temporal dynamics, which is our objective in this study, can therefore uncover the regulators of dynamic biological processes. Doing so can also help generate testable hypotheses about the roles of orphan genes that exhibit similar expression patterns to ones that are better understood. 

The complex, nonlinear time dynamics of gene expression pose a significant challenge for clustering and network analysis in genomics. Groups of interacting genes may be expressed with time lags or inverted patterns \citep{qian2001beyond} due to delayed activation of underlying transcription factors, making it difficult to measure the ``similarity'' in two expression profiles. Ordinary differential equations (ODEs) or discrete-time difference equations have been successfully used to model the nonlinear expressions of a small number of genes \citep{d1999linear, chen1999modeling, de2002modeling, bansal2006, polynikis2009comparing}. It is possible to derive similarity metrics for the time dynamics of two genes from such ODEs, thus enabling putative identification of co-regulated genes and the reconstruction of regulatory networks \citep{Farina2007, Farina2008, wu2019ODE}. In particular, the approach proposed in \citet{Farina2007} allows explicit modeling of lead-lag as well as contemporaneous associations between gene expression trajectories. We hence use it as the basis of the similarity calculations in our proposed clustering framework.

The high dimensionality (number of genes) and small sample sizes (number of time points) of time-course gene expression datasets pose another obstacle to identifying genes with similar expression dynamics. Due to the size of these datasets, the number of gene pairs receiving high similarity scores by any method can be overwhelmingly large. High similarity scores are typically validated for biological relevance using annotations provided by extensive public and commercial curated databases that assign genes to functional groups. For instance, Gene Ontology (GO) annotations are keywords that describe a gene's molecular function, role in a biological process, or cellular localization \citep{ashburner2000gene}. Other curated databases include KEGG \citep{kanehisa2000kegg}, Reactome \citep{fabregat2018reactome}, BioCyc \citep{karp2019biocyc}, and STRING \citep{szklarczyk2019string}. To ease the burden of manually validating a potentially vast number of gene-gene associations, we propose a Bayesian clustering technique that uses annotations as prior information to automatically validate these associations. Incorporating such information into a clustering method can encourage gene pairs with known biological associations to receive higher similarity scores, while filtering away those known to be unrelated. This also allows for knowledge gleaned from gene expression time series data to be contrasted with other knowledge bases; for instance, two genes with highly similar temporal expression patterns may not have been considered associated in previous cross-sectional (single time point) studies on which annotations are based, or vice versa.

There exist in the literature a few approaches to integrating biological knowledge with statistical measures of genetic association. One line of research considers Bayesian methods that use external data sources to determine prior distributions over genes or proteins that influence a biological response \citep{li2010bayesian, stingo2011, Hill2012, Lo2012, peng2013integrative}. Other studies develop biologically-informed regularization terms in graph-regularized methods for reconstructing gene networks  \citep{zhang2013molecular, li2015gene}. In another work, \citet{Nepomuceno2015} propose an algorithm for biclustering gene expression data using gene ontology annotations. However, less attention has been given to using both data and prior biological knowledge to identify and model dominant patterns in the complex temporal dynamics of gene expression, e.g. with ODEs.

\begin{figure}[H]
\centering\includegraphics[width=.8\textwidth]{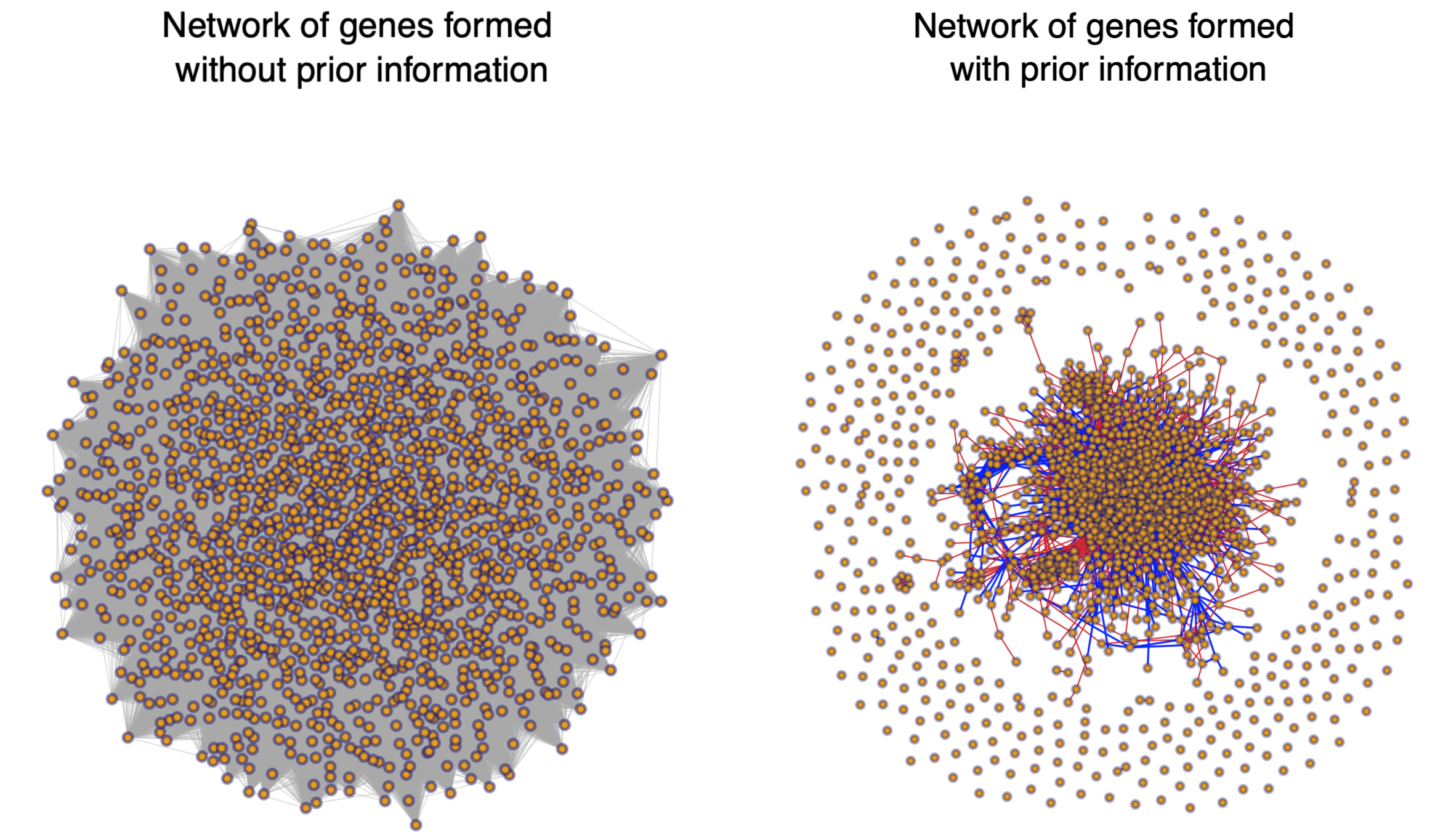}	
\caption{\textit{Networks of 1735 genes profiled in a time-course gene expression dataset collected by \cite{schlamp2020}. Vertices represent genes and edges connect two genes if their lead-lag $R^2$ exceeds 0.9. Left: lead-lag $R^2$ is computed using ordinary least squares regression according to \cite{Farina2008}, without any external biological information. All 1735 genes form a single connected component (599,896 edges). Right: lead-lag $R^2$ is computed using our proposed Bayesian approach, which leverages external sources of biological information about gene-gene relationships. Red edges (11,380 edges) connect genes known to be associated. Blue edges (2830 edges) connect genes whose relationship is unknown but is supported by the data.}}
\label{fig:networks}
\end{figure}

 Our technical contribution in this work is a Bayesian method for constructing biologically-meaningful clusters and networks of genes from time-course expression data, using a new similarity measure between two genes called the \textit{Bayesian lead-lag $R^2$} ($\LL R^2$). The Bayesian $\LL R^2$ is derived from ODE models of temporal gene expression, and is based on associations in both the time-course data and prior biological annotations. The balance between data and prior information is controlled by data-driven hyperparameters, making our approach an empirical Bayes method. As indicated by the name, the Bayesian $\LL R^2$ is based on the familiar $R^2$ statistic (the coefficient of determination) and is simple and fast to compute for all $\binom{N}{2}$ gene pairs, where $N$ is the number of genes under study. Importantly, external biological information regularizes the set of significant gene-gene associations found within a time-course dataset. In Figure \ref{fig:networks}, for instance, we present a network of 1735 genes in \textit{Drosophila melanogaster} (fruit fly) constructed both without external information, using an ordinary least-squares version of the $\LL R^2$ proposed by \cite{Farina2008}, and with external information, using our proposed Bayesian $\LL R^2$; the latter is a noticeably sparsified revision of the former, and retains only edges connecting genes with either known or highly plausible biological relationships.
 
The remainder of this paper is organized as follows. Section \ref{sec:geneExprModel} describes the ODE model of temporal gene expression that we adopt. Section \ref{sec:eBayes} details our empirical Bayes method for fitting the ODE model and obtaining our proposed $\LL R^2$ metric. Section \ref{sec:results} demonstrates the application of our method to real gene expression data collected by \citet{schlamp2020}. We recover a tradeoff between immune response and metabolism that has been observed in several studies and present  examples of biologically-meaningful gene clusters identified with the Bayesian $\LL R^2$. We also discuss the method's potential to conjecture new data-driven hypotheses about gene-gene interactions. Sections \ref{sec:immMet} and Appendices \ref{appsec:figures}-\ref{appsec:tables} compare the Bayesian $\LL R^2$ to its non-Bayesian counterpart as well as the commonly used Pearson correlation between genes. Proofs and additional examples are provided in the Appendix.

% TeX root = ../main.tex

\section{Dynamic models of gene expression}\label{sec:geneExprModel}

\subsection{Model derivation}\label{sec:modelDeriv}

We consider an ODE model of gene expression proposed by \citet{Farina2007}. Let $m_A(t)$ denote the expression of some gene $A$ at time $t$, measured for instance as the $\log_2$-fold change in mRNA levels relative to time 0. The model assumes the rate of change in gene $A$'s expression is given by some regulatory signal $p(t)$:
$$\frac{\dv m_A(t)}{\dv t}=p(t)-\kappa_A m_A(t)+\eta_A,$$
where $\kappa_A$ denotes the mRNA decay rate for gene $A$, and $\eta_A$ denotes natural and experimental noise. This model can be naturally extended to consider two genes $A$ and $B$ that might be associated with one another, i.e. that are governed by the same underlying $p(t)$, yielding a pair of coupled differential equations:
\begin{align}
\frac{\dv m_A(t)}{\dv t} &= \alpha_A p(t) + \beta_A - \kappa_A m_A(t) + \eta_A, \label{eq:dmA}\\ 
\frac{\dv m_B(t)}{\dv t} &= \alpha_B p(t) + \beta_B - \kappa_B m_B(t) + \eta_B. \label{eq:dmB}
\end{align}
The common signal $p(t)$ accounts for the effect of one or more transcription factors that potentially regulate both genes $A$ and $B$. The coefficients $\alpha_A$ and $\alpha_B$ measure the strength of $p(t)$ in the expression patterns of genes $A$ and $B$, respectively. $\beta_A$ and $\beta_B$ are affine coefficients allowing $m_A(t)$ and $m_B(t)$ to exhibit linear time trends.

We now obtain a model of gene $A$'s expression in terms of gene $B$'s expression by first rearranging (\ref{eq:dmB}) to isolate $p(t)$:
\begin{align}
p(t) = \frac{1}{\alpha_B}\left(\frac{\dv m_B(t)}{\dv t}-\beta_B+\kappa_B m_B(t)-\eta_B\right). \label{eq:pt}
\end{align}
Substituting (\ref{eq:pt}) into (\ref{eq:dmA}) yields
\begin{align*}
\frac{\dv m_A(t)}{\dv t}=\frac{\alpha_A}{\alpha_B}\frac{\dv m_B(t)}{\dv t} + \frac{\alpha_A\kappa_B}{\alpha_B}m_B(t) - \kappa_A m_A(t) + \beta_A-\frac{\alpha_A\beta_B}{\alpha_B} + \eta_A-\frac{\alpha_A\eta_B}{\alpha_B}.
\end{align*}
Integrating from $0$ to $t$, we obtain:
\begin{align*}
m_A(t) &= \frac{\alpha_A}{\alpha_B}m_B(t) + \frac{\alpha_A\kappa_B}{\alpha_B}\int_0^tm_B(s)\dv s - \kappa_A\int_0^t m_A(s)\dv s + \left(\beta_A-\frac{\alpha_A\beta_B}{\alpha_B}\right)t \\
&\hspace{1cm} + \int_0^t\left(\eta_A-\frac{\alpha_A\eta_B}{\alpha_B}\right)\dv s + c \\
&= c_1m_B(t) + c_2\int_0^tm_B(s)\dv s + c_3\int_0^t m_A(s)\dv s + c_4 t + c_5, \numberthis\label{eq:mA}
 \end{align*}
where $c_1 = \frac{\alpha_A}{\alpha_B}$,\ $c_2 =\frac{\alpha_A\kappa_B}{\alpha_B}$,\ $c_3 = -\kappa_A$,\ $c_4 = \beta_A-\frac{\alpha_A\beta_B}{\alpha_B}$, and $c_5 = \int_0^t\left(\eta_A-\frac{\alpha_A\eta_B}{\alpha_B}\right)\dv s+c$. 

Observe that (\ref{eq:mA}) is linear in the parameters $c_k$. Thus, given measurements $\{m_A(t_1),...,m_A(t_n)\}$ and $\{m_B(t_1),...,m_B(t_n)\}$ of the expression levels of genes $A$ and $B$ at time points $t_1,...,t_n$, we can express (\ref{eq:mA}) as the linear model $\mathbf{Y}=\mathbf{X}\bg\beta+\bg\varepsilon$, where
\begin{align}
\mathbf{Y} = \begin{bmatrix}m_A(t_1)\\[.5em] m_A(t_2) \\[.25em] ...\\[.25em] m_A(t_n)\end{bmatrix}, \hspace{.2cm}
\mathbf{X}=\begin{bmatrix} 
m_B(t_1) & \int_0^{t_1} m_B(s)\dv s & \int_0^{t_1}m_A(s)\dv s & t_1 & 1 \\[.5em]
m_B(t_2) & \int_0^{t_2} m_B(s)\dv s~ & \int_0^{t_2}m_A(s)\dv s & t_2 & 1 \\[.25em]
... & ... & ... & ... & ...\\[.25em]
m_B(t_n) & \int_0^{t_n} m_B(s)\dv s & \int_0^{t_n}m_A(s)\dv s & t_n & 1 \\ 	
\end{bmatrix}, \hspace{.2cm}\bg\beta =\begin{bmatrix}
	c_1 \\ c_2 \\ c_3\\ c_4\\ c_5
\end{bmatrix}, \label{eq:responseDesign}
\end{align}
with the standard assumption that $\bg{\varepsilon}\sim N(\mathbf 0,\sigma^2\mathbf{I}_n)$, where $\mathbf I_n$ denotes the $n\times n$ identity matrix. Although only samples from the functions $m_A(t)$ and $m_B(t)$ are given, we can estimate the integral entries of the second and third columns of $\mathbf{X}$ by numerically integrating spline or polynomial interpolants fitted to these samples.

In fitting the model (\ref{eq:mA}) to the expression data of genes $A$ and $B$, we obtain the ordinary least-squares (OLS) estimator $\bg{\hat\beta}=(\mathbf{X}^T\mathbf{X})^{-1}\mathbf{X}^T\mathbf{Y}$. The amount of variation in gene $A$'s expression that is captured by the estimated linear model is expressed as the familiar $R^2$ value: $R^2 =\|\mathbf{X}\bg{\hat\beta}-\overline{Y}\1_n\|^2 / \|\mathbf{Y}-\overline{Y}\1_n\|^2$, where $\overline Y=\frac{1}{n}\mathbf Y^T\1_n$ and $\1_n$ denotes the $n\times 1$ vector of ones. Adopting the terminology in \citet{Farina2008}, we refer to this $R^2$ as the \textit{lead-lag $R^2$ between genes $A$ and $B$}. A high lead-lag $R^2$ may indicate that the two genes are co-regulated in time by common transcription factors, or are at least associated with one another in some way. The term ``lead-lag'' comes from the lead-lag compensator in control theory. In this context, a ``lead-lag relationship'' between genes refers to the presence of a common regulatory signal (input) that, in conjunction with the process of mRNA decay, modulates the expression  of genes with the same biological function (output).

\subsection{Motivating the Bayesian lead-lag $R^2$}
\label{sec:challengesAndContributions}

Our primary contribution in this work is a biologically-informed method for clustering genes based on their temporal dynamics. Clustering involves measuring the similarity between two objects, which can also be thought of as defining an edge between two nodes in an undirected network. Our similarity measure is a Bayesian version of the lead-lag $R^2$ that uses both temporal expression data for genes $A$ and $B$ as well as a prior indication of whether they are biologically associated.

\begin{figure}[H]
\includegraphics[width=\textwidth]{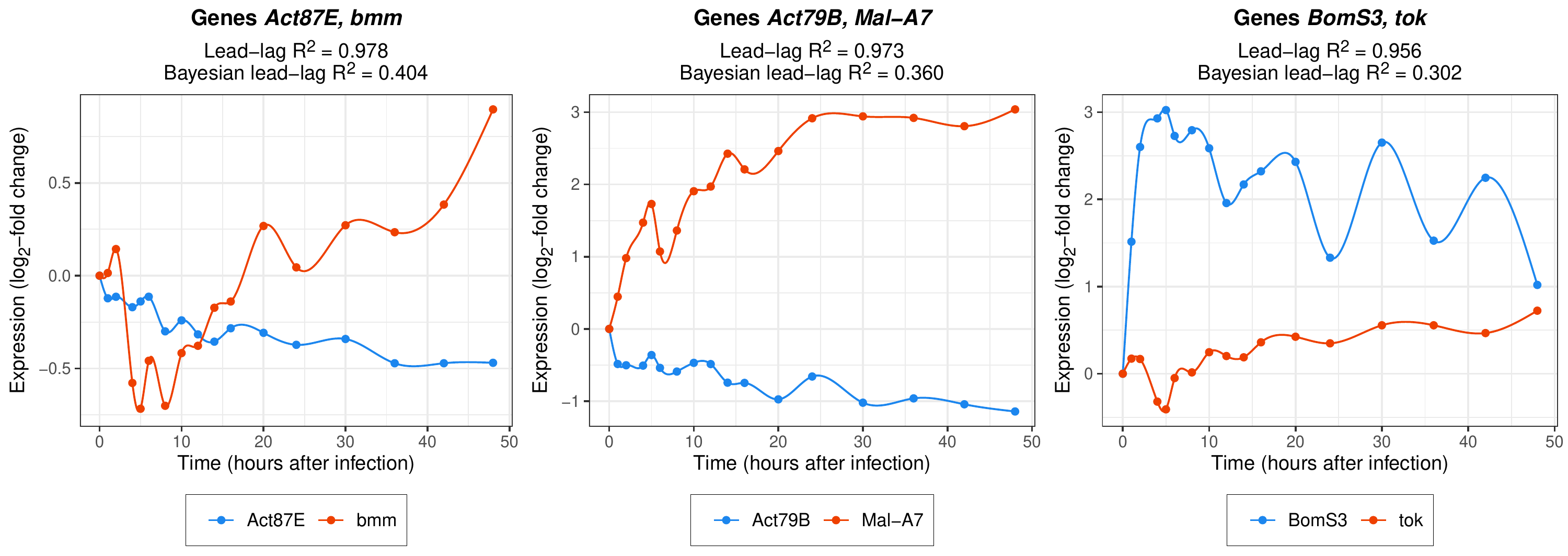}
%\end{figure}
%\begin{figure}[H]
\caption{\textit{Plots of the temporal expression profiles of three gene pairs for which the lead-lag $R^2$ is spuriously high. Spline interpolants were fit through the observed points, which are indicated by solid dots. Functional annotations for these genes in Flybase \citep{larkin2020flybase} do not suggest a clear link within each pair. By contrast, the lower Bayesian lead-lag $R^2$ values more accurately reflect the degree of these associations.}}
\label{fig:inflatedLLR2}
\end{figure}

We can motivate the Bayesian lead-lag $R^2$ via Figure \ref{fig:inflatedLLR2}, which shows examples of \textit{false positive} gene pairs: genes that have a spuriously high lead-lag $R^2$, but do not have similar expression patterns nor a biological relationship. The data comes from an experiment on fruit flies by \citet{schlamp2020} that aimed to profile the dynamics of genes involved in or affected by immune response immediately following an infection. More details on this dataset can be found in Section \ref{sec:dataset}.

Spuriously high lead-lag $R^2$ values are likely to arise in large datasets. For example, if gene $A$'s expression levels increase or decrease monotonically with time, the response vector $\mathbf Y$ in (\ref{eq:responseDesign}) will be highly correlated with the time integrals and time points in the third and fourth columns of $\mathbf X$. The lead-lag $R^2$ between genes $A$ and $B$ will be large, but not because the genes are associated either in time or biologically.

\begin{figure}[H]
\centering\includegraphics[width=\textwidth]{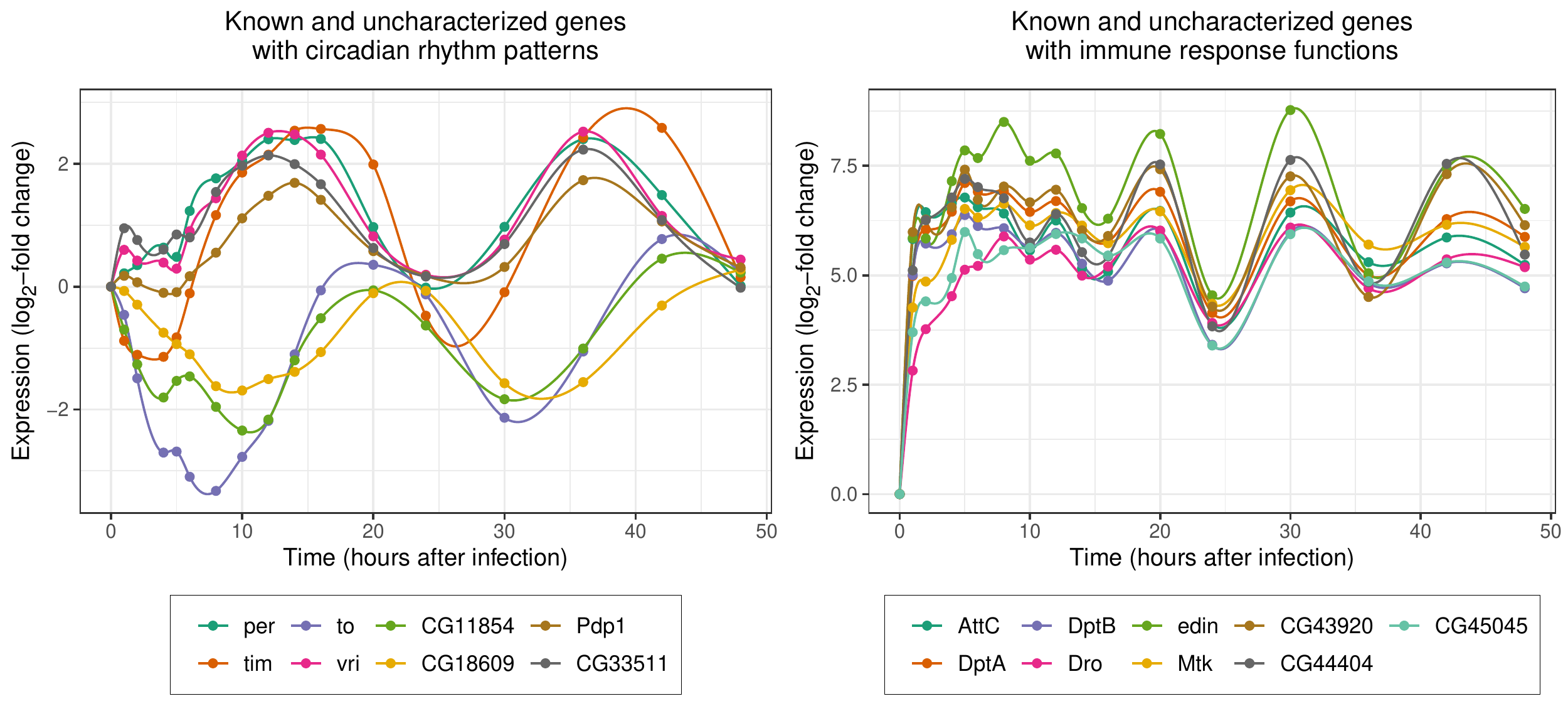}
%\end{figure}
%\begin{figure}[H]
\caption{\textit{Left: The uncharacterized gene \textit{CG33511} exhibits similar time dynamics to known circadian rhythm genes with 24-hour cyclic temporal expressions. Right: Uncharacterized genes \textit{CG43920} and \textit{CG45045} exhibit similar temporal expression patterns to known immune response genes, which are up-regulated in response to infection.}}
\label{fig:circadianImmune} 
\end{figure}

In contrast to gene pairs of the kind shown in Figure \ref{fig:inflatedLLR2}, we can consider Figure \ref{fig:circadianImmune}, which displays genes from two well-studied functional groups, known as \textit{pathways}: circadian rhythms and immune response. Within each group, we expect to see high pairwise lead-lag $R^2$ values (\textit{true positives}).

Incorporating pathway membership or protein-protein interaction networks into the lead-lag $R^2$ enables us to encourage genes in the same pathways to receive higher pairwise similarity scores, thus separating true positives from false positives. Importantly, we can also suggest possible pathways for previously uncharacterized genes. In our method, external biological information is used to determine the locations of normal prior distributions on the parameters $c_1,...,c_5$ in the model (\ref{eq:mA}). Upon obtaining the posterior estimates of these parameters, we recompute the lead-lag $R^2$ to obtain the \textit{Bayesian lead-lag $R^2$} ($\LL R^2$). In doing so, we observe a desirable shrinkage effect in the distribution of the Bayesian lead-lag $R^2$ values that pares down the number of significant associations. We next detail the hierarchical model and its hyperparameters in Section \ref{sec:eBayes}.
% TeX root = ../main.tex

\section{Empirical Bayes methodology}\label{sec:eBayes}

In this section, we propose our empirical Bayes approach to deriving biologically-informed similarity metrics between genes for clustering and network analysis. The components of our method are: 1) encoding external biological information into a prior adjacency matrix, 2) defining a normal-inverse gamma prior, specifically Zellner's $g$-prior, on the parameters of the ODE-based model of gene expression \eqref{eq:mA}, 3) optimally selecting the hyperparameter $g$ in Zellner's $g$-prior, and 4) calculating the Bayesian lead-lag $R^2$. Note that parts 2-4 lead to the computation of the Bayesian lead-lag $R^2$ for a single gene pair; a summary of the full algorithm for all pairs is provided in Appendix \ref{appsec:algo}.

\subsection{Part 1: Leveraging biological priors}\label{sec:bioPriors}

There exist numerous databases that extensively document known or predicted interactions between genes as well as their functional roles. For instance, Gene Ontology (GO) terms are keywords that describe a gene's molecular function, role in a biological process (e.g., ``lipid metabolism"), or cellular localization (e.g., ``nucleus") \citep{ashburner2000gene}. Semantic similarity methods, such as the R package \texttt{GOSemSim} \citep{yu2010gosemsim}, have been developed to determine how related genes are based on their associated GO terms. Other curated databases that similarly assign genes to pathways include KEGG \citep{kanehisa2000kegg}, Reactome \citep{fabregat2018reactome}, and BioCyc \citep{karp2019biocyc}. The STRING database \citep{szklarczyk2019string} aggregates multiple sources of information to generate a more holistic measurement of the association between two genes. For each pair of genes in an organism, STRING provides a score between 0 and 1 indicating how likely the two genes' encoded proteins are to interact physically based on experimental evidence, belong to the same pathways according to the aforementioned databases, be conserved across species, or be mentioned in the same studies. 

Regardless of which sources of external biological information one employs, the first step of our method is to encode this information into a matrix $\mathbf W$ of size $N\times N$, where $N$ is the total number of genes under study. The entries of $\mathbf W$ are:
\begin{align*}
\mathbf W_{ij} = 
\begin{cases}1 &\text{if prior evidence suggests that the } i^\tth \text{ and } j^\tth \text{ genes  are associated} \\
\text{NA} &\text{if the relationship between the } i^\tth \text{ and } j^\tth \text{ genes is unknown} \\
0 &\text{if the } i^\tth \text{ and } j^\tth \text{ genes are unlikely to be associated.}	
\end{cases}
\end{align*}

Intuitively, $\mathbf W$ can be viewed as an adjacency matrix for a network that reflects the known relationships amongst the genes. Our proposed method uses this network as an informed basis for the network constructed from the time-course gene expression data itself. Importantly, the data can indicate what the ``NA" entries of $\mathbf W$ ought to be, as well as confirm (or refute) the ``0'' or ``1'' entries.

In the event that $\mathbf W$ consists largely of ``NA'' entries, one can also use time-course gene expression data from previous studies to fill in some of them. Such datasets often consist of multiple replicates of the expression measurements, possibly gathered under the same experimental conditions to account for sampling variation, or different conditions to assess the effect of a treatment. 

\subsection{Part 2: The normal-inverse gamma model and Zellner's $g$-prior} \label{sec:normalInverseGamma}

Recall from Section \ref{sec:modelDeriv} that given measurements $\{m_A(t_1),...,m_A(t_n)\}$ and $\{m_B(t_1),...,m_B(t_n)\}$ of the expressions of two genes $A$ and $B$ at times $t_1,...,t_n$, the model we aim to fit is
$$m_A(t)=c_1m_B(t)+c_2\int_0^t m_B(s)\dv s + c_3\int_0^t m_A(s)\dv s + c_4t + c_5,$$
where $c_1,...,c_5$ are unknown parameters, and the integrals $\int_0^tm_A(s)\dv s$ and $\int_0^t m_B(s)\dv s$ are estimated by numerically integrating spline interpolants of the given data.  As seen in (\ref{eq:responseDesign}), this model can be represented in matrix form as $\mathbf Y=\mathbf X\bg\beta+\bg\varepsilon$, where $\bg\varepsilon\sim N(\mathbf 0,\sigma^2\mathbf I_n)$.

Since $c_1$ and $c_2$ link the expressions of genes $A$ and $B$, we intuitively ought to place priors of non-zero mean on these two parameters if $\mathbf W$ indicates that the genes are associated. To do this, we adopt the normal-inverse gamma model for $\bg\beta$ and $\sigma^2$, which is used frequently in Bayesian regression and allows for flexible modeling of the prior mean and covariance matrix of $\bg\beta$. The normal-inverse gamma model specifies $\bg\beta|\sigma^2 \sim N\left(\bg\beta_0, \sigma^2\mathbf{V}_0\right)$ and $\sigma^2 \sim \Gamma^{-1}(a, b)$ where $\bg\beta_0\in\R^{p\times 1}$, $\mathbf{V}_0\in\R^{p\times p}$ is a positive semidefinite matrix, and $a,b>0$. It is then said that $(\bg\beta,\sigma^2)$ jointly follow a normal-inverse gamma distribution with parameters $(\bg{\beta}_0,\mathbf{V}_0,a,b)$. This is a conjugate prior, so the posterior distribution of $(\bg\beta,\sigma^2)$ is also normal-inverse gamma with parameters $(\bg\beta_*, \mathbf{V}_*,a_*,b_*)$ defined as 
\begin{gather}
\bg\beta_* = \left(\mathbf{V}_0^{-1} + \mathbf{X}^T\mathbf{X}\right)^{-1}\left(\mathbf{V}_0^{-1}\bg\beta_0 + \mathbf{X}^T\mathbf{Y}\right), \label{eq:betaPostMean}\\
\mathbf{V}_* = \left(\mathbf{V}_0^{-1} + \mathbf{X}^T\mathbf{X}\right)^{-1}, \label{eq:betaPostCov}\\
a_* = a + \frac{n}{2} \label{eq:aPost}, \\
b_* = b + \frac{1}{2}\left(\bg{\beta}_0^T\mathbf{V}_0^{-1}\bg\beta_0 + \mathbf{Y}^T\mathbf{Y} - \bg\beta_*^T\mathbf{V}_*^{-1}\bg\beta_*\right). \label{eq:bPost}
\end{gather}
That is, the conditional posterior  of $\bg\beta$ given $\sigma^2$ and the posterior  of $\sigma^2$ are $\bg\beta|\sigma^2,\mathbf{Y}\sim N(\bg\beta_*,\sigma^2\mathbf{V}_*)$ and $\sigma^2|\mathbf{Y} \sim \Gamma^{-1}(a_*,b_*)$. The posterior mean $\bg\beta_*$ can be used as the estimated regression parameters. This estimator has the desirable properties of consistency and asymptotic efficiency in large samples and admissibility in finite samples  \citep{giles1979mean}. 

The hyperparameters $\bg\beta_0$ and $\mathbf V_0$ are of particular interest as they allow us to incorporate biological information into our model. In defining $\bg\beta_0$, recall that we wish to place priors with non-zero mean on the parameters $c_1$ and $c_2$ when external sources suggest that genes $A$ and $B$ are co-regulated or at least associated. We noted in Section \ref{sec:modelDeriv} that $c_1$ represents the ratio $\alpha_A/\alpha_B$, where $\alpha_A$ and $\alpha_B$ denote the strength of a common regulatory signal in the first-order dynamics of the two genes. If the genes are associated, it is reasonable to believe \textit{a priori} that $\alpha_A=\alpha_B$, implying $c_1=1$. Then we could also say \textit{a priori} that $c_2=1$, since $c_2$ represents a parameter that is proportional to $c_1$. When prior information about genes $A$ and $B$ is lacking or suggests that they are unrelated, a prior mean of zero for $c_1$ and $c_2$ is appropriate. Supposing genes $A$ and $B$ are the $i^\tth$ and $j^\tth$ genes in the dataset, we can thus set the prior mean $\bg\beta_0$ as follows:
\begin{align*}
\boldsymbol\beta_0 = \begin{cases} [1,1,0,0,0]^T  &\text{if } \mathbf W_{ij}=1 \\ [0,0,0,0,0]^T  & \text{if } \mathbf W_{ij}=0 \text{ or NA.} \end{cases}
\end{align*}

As for the prior covariance matrix $\sigma^2 \mathbf V_0$ of $\bg\beta$, we first note that for the linear model $\mathbf Y=\mathbf X\bg\beta+\bg\varepsilon$ where $\bg\varepsilon$ has the covariance matrix $\sigma^2\mathbf I_n$, the least-squares estimator $\bg{\hat\beta}_{\text{OLS}}=(\mathbf{X}^T\mathbf{X})^{-1}\mathbf{X}^T\mathbf{Y}$ has the covariance matrix $\sigma^2(\mathbf{X}^T\mathbf{X})^{-1}$. A popular choice for $\mathbf V_0$ is therefore $g(\mathbf{X}^T\mathbf{X})^{-1}$, where $g>0$. This choice of $\mathbf V_0$ yields a particularly tractable special case of the normal-inverse gamma model known as \textit{Zellner's $g$-prior} \citep{zellner1986}. Substituting this choice of $\mathbf V_0$ into the posterior mean $\bg\beta_*$ in (\ref{eq:betaPostMean}) and covariance matrix $\mathbf V_*$ in (\ref{eq:betaPostCov}), we obtain
\begin{gather}
\bg\beta_* = \frac{1}{1+g}\boldsymbol\beta_0 + \frac{g}{1+g}\bg{\hat\beta}_{\text{OLS}}, \label{eq:betaPostGPrior}\\
\mathbf V_* = \frac{g}{1+g}(\mathbf{X}^T\mathbf{X})^{-1}. \label{eq:betaPostCovGPrior}
\end{gather}
(\ref{eq:betaPostGPrior}) reveals that under Zellner's $g$-prior, the posterior mean $\bg\beta_*$ is a convex combination of the prior mean $\bg\beta_0$ and the least-squares estimator $\bg{\hat\beta}_\text{OLS}$. The parameter $g$ balances the weights placed on external information encoded by $\bg\beta_0$ and on the data used to compute $\bg{\hat\beta}_\text{OLS}$, so the selection of $g$ is an important component of our modeling strategy. We next describe our data-driven approach to choosing it.

\subsection{Part 3: Optimal selection of $g$ in Zellner's $g$-prior} \label{sec:optimalG}

Several methods for choosing $g$ in Zellner's $g$-prior have been proposed previously. For instance, \citet{GeorgeFoster2000} discuss an empirical Bayes method in which one selects the value of $g$ that maximizes the marginal likelihood of $\mathbf Y$. \citet{Liang2008} provide a closed form expression for this maximizing value of $g$ that is nearly identical to the $F$-statistic for testing the hypothesis that $\bg\beta=\mathbf 0$. They show that this maximum marginal likelihood approach has the desirable property that the Bayes factor for comparing the full model to the null (intercept-only) model diverges as the $R^2$ approaches 1. For our application, one concern is that the $F$-statistic defining this particular estimate of $g$ is likely to be overwhelmingly large for many gene pairs. If $g$ is large, then $\bg\beta_*$ will be very close to $\bg{\hat\beta}_{\text{OLS}}$ according to (\ref{eq:betaPostGPrior}). As a result, any biological evidence of association captured by $\bg\beta_0$ will have a negligible impact on the model. 

A fully Bayesian approach to selecting $g$ involves placing a prior distribution on $g$ that is then integrated out when defining the prior on $\bg\beta$. This method is motivated by \citet{Zellner1980}, who propose placing a Cauchy prior on $\bg\beta$ to derive a closed-form posterior odds ratio for hypothesis testing purposes. These priors can be represented as a mixture of $g$-priors with an inverse gamma prior on $g$, although a closed-form posterior estimate of $g$ is unavailable. \citet{cui2008} and \citet{Liang2008} alternatively consider a class of priors of the form $\pi(g) = \frac{a-2}{2}(1+g)^{a/2}$ for $a>2$, known as the hyper-$g$ priors. Under these priors, the posterior mean of $g$ can be expressed in closed form in terms of the Gaussian hypergeometric function. 

Because our application involves fitting regression models for potentially thousands of gene pairs, the computational cost of fully Bayesian methods for selecting $g$ requires us to consider alternative approaches. One idea is to select the value of $g$ that minimizes the sum of squared residuals $\|\mathbf Y-\mathbf{\hat Y}\|^2$, where $\mathbf{\hat Y}=\mathbf X\bg\beta_*$ is the vector of fitted values: 
\begin{align}
\mathbf{\hat Y}\;=\;\mathbf X\bg\beta_*\;=\;\mathbf X\left(\frac{1}{1+g}\bg\beta_0 +\frac{g}{1+g}\bg{\hat\beta}_{\text{OLS}}\right) \;=\; \frac{1}{1+g}\mathbf Y_0+\frac{g}{1+g}\mathbf{\hat Y}_{\text{OLS}}, \label{eq:YpostMean}
\end{align}
where $\mathbf Y_0=\mathbf X\bg\beta_0$ and $\mathbf{\hat Y}_\text{OLS}=\mathbf X\bg{\hat\beta}_{\text{OLS}}$. However, we found that there are no analytical solutions to $g_*=\argmin_{g>0}\|\mathbf Y-\mathbf{\hat Y}\|^2=0$. Instead, we can minimize Stein's unbiased risk estimate (SURE), which is an unbiased estimate of $\|\mathbf{\hat Y}-\mathbf X\bg\beta\|^2$. Below is a rephrased version of Theorem 8.7 in \citet{Wells2018}, which defines SURE for the general problem of estimating $\E(\mathbf Y)$ using a linear estimator of the form $\mathbf{\hat Y} = \mathbf{a+SY}$. This theorem statement differs from its original version in that we have rewritten the divergence of $\mathbf X\bg{\hat \beta}$ with respect to $\mathbf Y$ using the generalized degrees of freedom developed by \citet{efron2004}.

\begin{theorem}[SURE for linear models]\label{thm:sure}
Let $\mathbf Y\sim N(\mathbf X\bg\beta,\sigma^2\mathbf I_n)$, where the dimensions of $\mathbf X$ are $n\times p$, and let $\bg{\hat\beta}=\bg{\hat\beta}(\mathbf Y)$ be a weakly differentiable function of the least squares estimator $\bg{\hat\beta}_{\text{OLS}}$ such that $\mathbf{\hat Y}=\mathbf X\bg{\hat\beta}$ can be written in the form $\mathbf{\hat Y}=\mathbf a+\mathbf{SY}$ for some vector $\mathbf a$ and matrix $\mathbf S$. Let $\hat\sigma^2 = \|\mathbf Y-\mathbf X\bg{\hat\beta}_\text{OLS}\|^2/(n-p)$. Then,
\begin{align}
\delta_0(\mathbf Y) = \|\mathbf Y - \mathbf X\bg{\hat\beta}\|^2 + \left(2\:\text{Tr}(\mathbf S)-n\right)\hat\sigma^2 \label{eq:sure0}
\end{align}
is an unbiased estimator of $\|\mathbf{\hat Y}-\mathbf X\bg\beta\|^2$. 
\end{theorem}

From (\ref{eq:YpostMean}), observe that we can write $\mathbf{\hat Y}$ as $ \frac{1}{1+g}\mathbf Y_0 +  \frac{g}{1+g}\mathbf{HY}$,
where $\mathbf H= \mathbf X(\mathbf{X}^T\mathbf{X})^{-1}\mathbf{X}^T$. Therefore the matrix $\mathbf S$ in Theorem \ref{thm:sure} is $g\mathbf H/(1+g)$, whose trace is $gp/(1+g)$. SURE in (\ref{eq:sure0}) then becomes
\begin{align}
\delta_0(\mathbf Y)=\|\mathbf Y - \mathbf X\bg\beta_*\|^2 + \left(\frac{2gp}{1+g}-n\right)\hat\sigma^2, \label{eq:sure}
\end{align}
where we have also substituted the posterior mean $\bg\beta_*$ in (\ref{eq:betaPostGPrior}) for $\bg{\hat\beta}$.

We next present the value of $g$ that minimizes SURE.
\begin{theorem}[SURE minimization with respect to $g$]\label{thm:SUREoptg}
	The value of $g$ that minimizes SURE in (\ref{eq:sure}) is
$$g_* = \frac{\|\mathbf{\hat Y}_{\text{OLS}}-\mathbf Y_0\|^2}{p\hat\sigma^2} - 1
.$$
\end{theorem}

The proof of Theorem \ref{thm:SUREoptg} is provided in Appendix \ref{appsec:proofs} .

It is quite possible that $p\hat\sigma^2$ is small, resulting in a large value of $g_*$ (i.e., $g_*\gg 1$) in Theorem \ref{thm:SUREoptg}. In this case, $\bg\beta_*$ in (\ref{eq:betaPostGPrior}) will  be largely influenced by the data via $\bg{\hat\beta}_\text{OLS}$, rather than by prior information via $\bg\beta_0$. This is desirable when the relationship between the two genes is unknown (i.e. $\mathbf W_{ij}=\text{NA}$), but  not if the relationship is known to be unlikely (i.e. $\mathbf W_{ij}=0$). In the latter case, we  prefer to shrink the regression coefficients towards the prior mean $\bg\beta_0=\mathbf 0$ so as to yield a smaller lead-lag $R^2$ value. To address this, we set $g$ conditionally on $\mathbf W_{ij}$ as
\begin{align}
g = \begin{cases} g_*  &\text{if } \mathbf W_{ij}=\text{NA or } 1 \\
 1 &\text{if } \mathbf W_{ij}=0,	
 \end{cases} \label{eq:gCond}
\end{align}
\noindent where $g_*$ is defined according to  Theorem \ref{thm:SUREoptg}. 

\subsection{Part 4: Computing the $R^2$ for Bayesian regression models} \label{sec:R2calculation}

Once a posterior estimate of the model coefficients (\ref{eq:betaPostGPrior})  has been obtained, with the parameter $g$ selected optimally, we can compute the Bayesian lead-lag $R^2$ between genes $A$ and $B$. 

Recall that for a linear model $\mathbf Y=\mathbf X\bg\beta +\bg\varepsilon$ where $\bg\beta$ is estimated by the least-squares estimator $\bg{\hat\beta}_{\text{OLS}} = (\mathbf X^T\mathbf X)^{-1}\mathbf X^T\mathbf Y$, the $R^2$ is defined as
\begin{align}
R^2 =\frac{\|\mathbf{X}\bg{\hat\beta}_{\text{OLS}}-\overline{Y}\1_n\|^2}{\|\mathbf{Y}-\overline{Y}\1_n\|^2}, \label{eq:R2ols}
\end{align}
where $\overline{Y} = (1/n)\mathbf Y^T\1_n$. In Bayesian regression, however, the standard decomposition of total sum of squares into residual and regression sums of squares no longer holds. Thus, when we replace $\bg{\hat\beta}_{\text{OLS}}$ with an estimator such as the posterior mean of $\bg\beta$, the formula (\ref{eq:R2ols}) can potentially yield $R^2 >1$. As a remedy to this issue, we compute the $R^2$ as the ratio of the variance of the model fit to itself plus the variance of the residuals. This ratio is within $[0,1]$ by construction, and is given by
\begin{align}
R^2 = \frac{\widehat{\var}(\mathbf X\bg \beta_*)}{\widehat\var(\mathbf X\bg\beta_*) + \widehat\var(\mathbf Y-\mathbf X\bg\beta_*)}, \label{eq:bayesLLR2}
\end{align}
where, for a vector $\mathbf Z=[z_1\;...\;z_n]^T$ with mean $\bar z$, we define $\widehat\var(\mathbf Z) = \frac{1}{n-1}\sum_{i=1}^n (z_i-\bar{z})^2 $. This calculation is based on the approach to computing $R^2$ for Bayesian regression models proposed in \citet{Gelman2018}. For our application, we refer to (\ref{eq:bayesLLR2}) as the \textit{Bayesian lead-lag} $R^2$ ($\LL R^2$).

\subsection{Clustering and empirical analysis with the Bayesian lead-lag $R^2$}\label{sec:clusteringMethod}

Given a dataset of $N$ genes whose expressions are measured at $n$ time points, our objective is to cluster the genes based on their temporal expression patterns. To do this, we compute a $N\times N$ similarity matrix $\mathbf S$ where $\mathbf S_{i,j}$ stores the Bayesian $\LL R^2$ in \eqref{eq:bayesLLR2} between the $i^\tth$ and $j^\tth$ genes. We then apply hierarchical clustering to the distance matrix $\mathbf J - \mathbf S$, where $\mathbf J$ is the $N\times N$ matrix of ones.

Note that the Bayesian $\LL R^2$ is asymmetric: $\LL R^2(i,j)\neq \LL R^2(j,i)$. Here, $\LL R^2(i,j)$ denotes the Bayesian $\LL R^2$ where we treat the $i^\tth$ gene as the response (gene $A$) in the model \eqref{eq:mA}. For our purpose of clustering a large set of genes for empirical analysis, we symmetrize the Bayesian $\LL R^2$ by setting:
$$\mathbf S_{i,j} = \max\left\{\LL R^2(i,j), \LL R^2(j,i)\right\}.$$

In practice, it is also common to use similarity measures such as the Bayesian $\LL R^2$ to produce a ranked list of gene-gene associations. To aid this procedure, we further propose two additional metrics that one could use in conjunction with the Bayesian $\LL R^2$. These metrics are derived from the following two sub-models of the original model \eqref{eq:mA}: 
\begin{align}
&m_{A}(t) = c_1m_B(t)+c_2\int_0^tm_B(s)\dv s +  c_5, &\text{(Sub-model 1)} \label{eq:subModel1}\\ 
&m_A(t) = c_3\int_0^t m_A(s)\dv s + c_4 t + c_5. &\text{(Sub-model 2)} \label{eq:subModel2}
\end{align}
The first sub-model describes the temporal expression of gene $A$ as a function only of the potentially co-regulated gene $B$. The second sub-model is ``autoregressive'' in the sense that it describes gene $A$'s expression only in terms of its own past and linear time trends. We again apply our Bayesian approach to fitting these two sub-models and compute new variants of the Bayesian $\LL R^2$ from each, denoted $\LL R^2_\other$ and $\LL R^2_\own$ respectively. The $\LL R^2_\other$ value indicates the amount of variation in gene $A$'s temporal expression that can be explained by the dynamics of an\textit{other} gene $B$. $\LL R^2_\own$ indicates the amount of variation in gene $A$ that is explained by its \textit{own} past, via its time integrals, and linear time trends. We can view $\LL R^2-\LL R^2_\own$ as the amount of \textit{additional} variation in gene $A$'s temporal dynamics that can be explained by considering gene $B$, on top of the variation captured by gene $A$'s own past via sub-model 2. Intuitively, if $\LL R^2$ is large, then a large value of $\LL R^2-\LL R^2_\own$ suggests that the $\LL R^2$ value is unlikely to mean a false positive relationship between the two genes. In Section \ref{sec:R2scatter}, we demonstrate empirically that using $\LL R^2_\other$ and $\LL R^2-\LL R^2_\own$ together can help identify gene pairs with highly similar time dynamics.

\subsection{Significance of the Bayesian lead-lag $R^2$}
One may further desire a notion of statistical significance for the Bayesian $\LL R^2$. One option is to simulate the posterior distribution of the $\LL R^2$ using draws from the posterior distribution of $\bg\beta$, as described in \citet{Gelman2018}. Recall from Section \ref{sec:normalInverseGamma} that the posterior distribution of $(\bg\beta,\sigma^2)$ is normal-inverse gamma with parameters $(\bg\beta_*, \mathbf V_*, a_*,b_*)$, defined respectively in (\ref{eq:betaPostGPrior}), (\ref{eq:betaPostCovGPrior}), (\ref{eq:aPost}), and (\ref{eq:bPost}). To draw samples from this posterior distribution, we can first sample $\sigma^2$ from the $\Gamma^{-1}(a_*,b_*)$ distribution, and then sample $\bg\beta$ from the $N(\bg\beta_*, g\sigma^2\mathbf V_*)$ distribution. In particular, if $\mathbf W_{ij}=1$, we may wish to simulate the posterior distribution of the Bayesian $\LL R^2$ under a null hypothesis of no relationship between genes $A$ and $B$. This can be reflected in the sampling procedure by calculating $\bg\beta_*$ with $\bg\beta_0$ set to $\mathbf 0$. 

Alternatively, one could use the Bayes factors presented in \citet{Liang2008} to corroborate the Bayesian lead-lag $R^2$. Let $\mc M_F$ denote the model (\ref{eq:mA}) of gene expression, which has an intercept and $p=4$ ``covariates'' with coefficients $c_1$ through $c_4$. Let $\mc M_N$ denote the null (intercept-only) model. Then the Bayes factor for comparing $\mc M_F$ to $\mc M_N$ is
\begin{align*}
\text{BF}(\mc M_F : \mc M_N) = (1+g)^{(n-p-1)/2}\frac{1}{\left(1+g(1-R^2)\right)^{(n-1)/2}}\:,
\end{align*}
where $R^2$ is the usual coefficient of determination in (\ref{eq:R2ols}). \citet{kass1995bayes} interpret a $\log_{10}(\text{BF}(\mc M_F :\mc M_N))$ value between 1 and 2 as ``strong'' evidence in favor of $\mc M_F$, or above 2 as ``decisive'' evidence.

In Appendix \ref{appsec:algo}, we give a summary of our methodology in the form of a generic algorithm that can be run on any time-course gene expression dataset.

\subsection{Possible modifications to prior hyperparameters}

In Section \ref{sec:normalInverseGamma}, we set the prior mean $\bg\beta_0$ of the parameters of the ODE model to $[1,1,0,0,0]^T$ when $\mathbf W_{ij}=1$, i.e. there is prior evidence suggesting genes $A$ and $B$ are associated. To make the method even more data-driven, one could alternatively set $\bg\beta_0=[\xi,\xi,0,0,0]^T$ in the $\mathbf W_{ij}=1$ case, where $\xi\neq 0$ is chosen adaptively from the data. The following theorem presents the values of $\xi$ and $g$ that simultaneously minimize SURE in (\ref{eq:sure}) in this setting. 

\begin{theorem}[SURE minimization with respect to $\xi$ and $g$]\label{thm:sureOptGXi}
Assume the entries of the least-squares estimator $\bg{\hat\beta}_{\text{OLS}}$ are all distinct and  the expression of gene $B$ is non-zero for at least one time point, i.e. $m_B(t_i)\neq 0$ for at least one $i$. Let $\bg\beta_0 = [\xi,\xi,0,0,0]^T$ in the case that $\mathbf W_{ij}=1$. Then the values of $\xi$ and $g$ that minimize SURE in (\ref{eq:sure}) are
\begin{align*}
\xi_* = \frac{\mathbf Y^T\mathbf X_{12}}{\|\mathbf X_{12}\|^2} \;\;\;\text{ and }\;\;\; g_* = \frac{\|\mathbf{\hat Y}_{\text{OLS}}\|^2\|\mathbf X_{12}\|^2 - (\mathbf Y^T\mathbf X_{12})^2}{\|\mathbf X_{12}\|^2p\hat\sigma^2} - 1,
\end{align*}
where $\mathbf X_{12}\in\R^{n\times 1}$ is the element-wise sum of the first two columns of $\mathbf X$.
\end{theorem}

The proof of Theorem \ref{thm:sureOptGXi} is provided in Appendix \ref{appsec:proofs}.

% TeX root = ../main.tex

\section{Results for the \textit{Drosophila melanogaster} dataset}\label{sec:results}

\subsection{Collecting a time-course gene expression dataset}\label{sec:dataset}

The expression of a gene is typically measured via RNA sequencing (RNA-seq) as the count of a particular type of messenger RNA found in a tissue sample. These counts can be normalized to library size and transformed using the limma-voom transformation \citep{law2014voom}. This transformation produces near-normally distributed gene expression measurements, making them more amenable to analysis with linear models such as those described in Section \ref{sec:modelDeriv}.

Our primary time-course gene expression dataset, introduced in \citet{schlamp2020}, profiles the expression dynamics of 12,657 genes in \textit{Drosophila melanogaster} (fruit fly) in response to an immune challenge. Immune responses were triggered in flies by injecting them  with commercial lipopolysaccharide, which contains peptidoglycan (cell wall molecules) derived from the bacterium \textit{E. coli}. Following injection, the flies were sampled via RNA-seq at 20 time points over five days. The data was normalized by the aforementioned limma-voom transformation and expressed as the $\log_2$-fold change with respect to the first time point, which was sampled pre-injection as a control. We focus on the first 17 time points, ranging from zero to 48 hours post-injection, as this is when most of the variation in expression occurs. 

Differential expression analysis is typically used to identify genes exhibiting significant expression changes, and thus reduce a set of thousands of genes into a smaller set meriting further study. In Appendix \ref{appsec:data}, we provide further details on how we use such methods to reduce our set of 12,657 genes into a set of 1735 genes of interest. We also describe therein how we define our prior adjacency matrix $\mathbf W$, introduced in Section \ref{sec:bioPriors}, using the STRING database.

\subsection{Small-scale case study: immunity and metabolism} \label{sec:immMet}

\begin{figure}[H]
\centering\includegraphics[width=10cm]{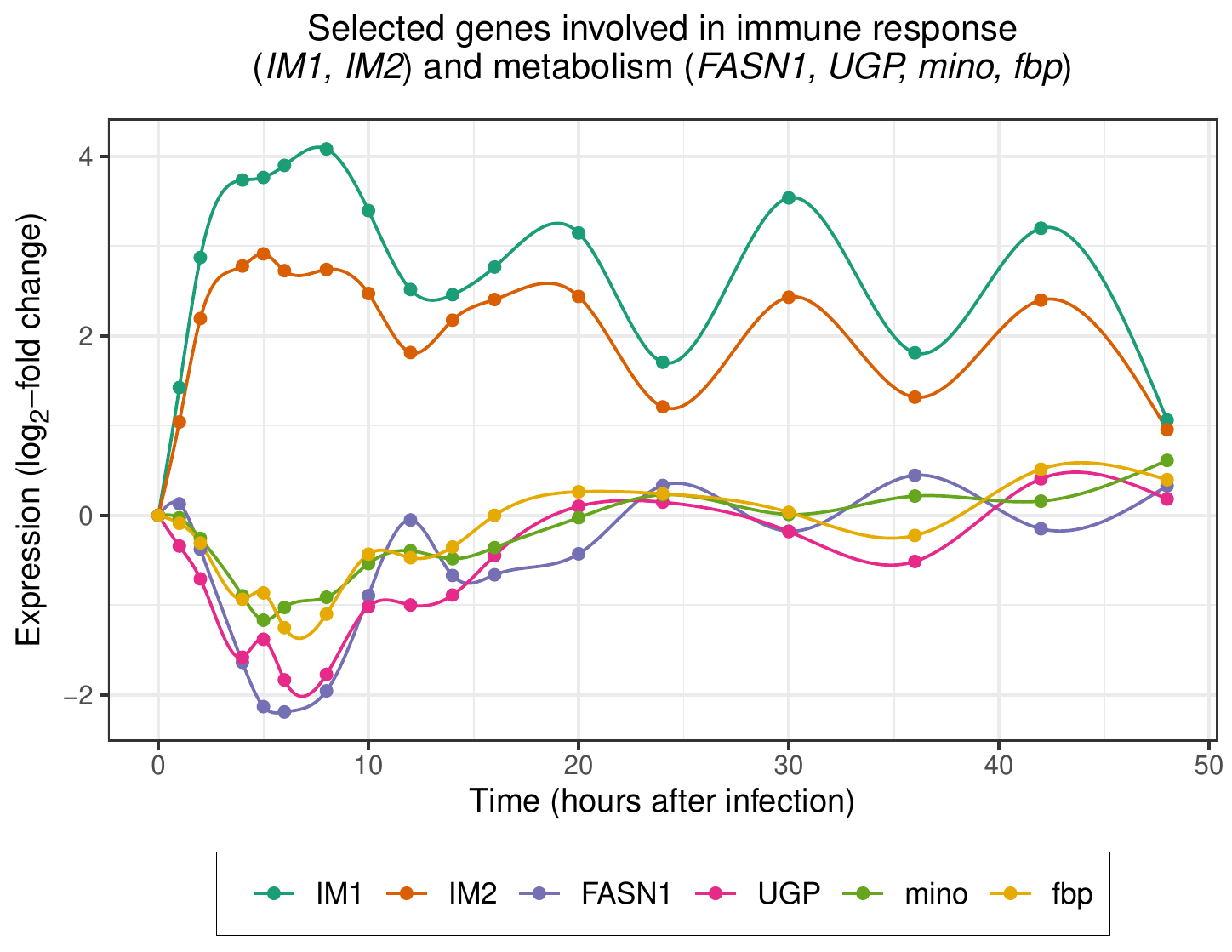}
%\end{figure}
%\begin{figure}[H]
\caption{\textit{Temporal expression patterns of selected genes involved in immune response (IM1, IM2) and metabolic processes (FASN1, UGP, mino, fbp). Upon infection, immune response genes are immediately up-regulated. At the same time, metabolic genes are down-regulated but return to pre-infection expression levels after 12 to 24 hours, which is before the immune response is resolved.}}
\label{fig:immMet}
\end{figure}

We first validate our methodology on a small set of genes whose behavior exhibits a known interplay between immunity and metabolism. \citet{schlamp2020} observe that exposure to bacterial peptidoglycan has an effect not only on the time dynamics of immune response, but also on the expression of genes involved in metabolism. In particular, some genes involved in immune response are up-regulated immediately following peptidoglycan injection, while other genes associated with metabolic processes are down-regulated more slowly. Interestingly, the metabolic genes return to their pre-infection levels of expression well before the immune response has subsided. This phenomenon can be observed in Figure \ref{fig:immMet}, which shows the expression patterns of two immune response genes (\textit{IM1, IM2}) and four metabolic process genes (\textit{FASN1, UGP, mino, fbp}). 

Tables \ref{tab:immMetW}-\ref{tab:NonBayesLLR2} show the prior adjacency matrix $\mathbf W$, the Bayesian $\LL R^2$ values, and the non-Bayesian $\LL R^2$ values (computed via ordinary least-squares regression) corresponding to these six genes. 

\begin{table}[H]
\small\centering
% -- Table 1 --
\begin{tabular}{|c|c|c!{\vrule width 1pt}c|c|c|c|}\hline
\setlength\tabcolsep{4.2pt}
\cellcolor{graycell}&  \cellcolor{graycell}{IM1} & \cellcolor{graycell}{IM2} & \cellcolor{graycell}{FASN1} & \cellcolor{graycell}{UGP} & \cellcolor{graycell}{mino} & \cellcolor{graycell}{fbp} \\ \hline
\cellcolor{graycell}{IM1} & - & \cellcolor{redcell}1 & NA & 0 & 0 & NA \\ \hline
\cellcolor{graycell}{IM2} & \cellcolor{redcell}1 & - & NA & 0 & 0 & NA\\ \noalign{\hrule height 1pt}%
\cellcolor{graycell}{FASN1} & NA & NA & - & NA & NA & NA\\ \hline
\cellcolor{graycell}{UGP} & 0 & 0 & NA & - & \cellcolor{redcell}1 & NA\\ \hline
\cellcolor{graycell}{mino} & 0 & 0 & NA & \cellcolor{redcell}1 & - & NA \\ \hline
\cellcolor{graycell}{fbp} & NA & NA & NA & NA & NA & - \\ \hline
\end{tabular}
\caption{\textit{Portion of the prior adjacency matrix $\mathbf W$ corresponding to the six genes in Figure \ref{fig:immMet}. Red cells indicate that there is prior evidence of association between the two genes. ``NA'' entries indicate that the association between the two genes is unavailable in the STRING database.}}\label{tab:immMetW}
\end{table}

\begin{table}[H]
\small\centering
% -- Table 2 --
\parbox{.48\linewidth}{
\setlength\tabcolsep{4pt}
\begin{tabular}{|c|c|c!{\vrule width 1pt}c|c|c|c|}\hline
\cellcolor{graycell}&  \cellcolor{graycell}{IM1} & \cellcolor{graycell}{IM2} & \cellcolor{graycell}{FASN1} & \cellcolor{graycell}{UGP} & \cellcolor{graycell}{mino} & \cellcolor{graycell}{fbp} \\ \hline
\cellcolor{graycell}{IM1} & - & \cellcolor{redcell}0.99 & \cellcolor{bluecell}0.82 & 0.30 & 0.40 & 0.68\\ \hline
\cellcolor{graycell}{IM2} & \cellcolor{redcell}0.99 & - & \cellcolor{bluecell}{0.80} & 0.30 & 0.39 & 0.66 \\ \noalign{\hrule height 1pt}%
\cellcolor{graycell}{FASN1} & \cellcolor{bluecell}0.82 & \cellcolor{bluecell}0.80 & - & \cellcolor{bluecell}0.83 & \cellcolor{bluecell}0.98 & \cellcolor{bluecell}0.82\\ \hline
\cellcolor{graycell}{UGP} & 0.30 & 0.30 & \cellcolor{bluecell}0.83 & - & \cellcolor{redcell}0.91 & \cellcolor{bluecell}0.99 \\ \hline
\cellcolor{graycell}{mino} & 0.40 & 0.39 & \cellcolor{bluecell}0.98 & \cellcolor{redcell}0.91 & - & \cellcolor{bluecell}0.90 \\ \hline
\cellcolor{graycell}{fbp} & 0.68 & 0.66 & \cellcolor{bluecell}0.82 & \cellcolor{bluecell}0.99 & \cellcolor{bluecell}0.90 & - \\ \hline
\end{tabular}\caption{\textit{Bayesian $\LL R^2$ values corresponding to each gene pair in Figure \ref{fig:immMet}. Colored cells mark values above 0.76, the 95$^\tth$ percentile of the empirical distribution of this metric for this dataset. Red cells are consistent with prior evidence of association (c.f. red cells in Table \ref{tab:immMetW}). Blue cells indicate potential associations that may not have been known previously. Overall, colored cells point to biologically-meaningful relationships within the immunity and metabolism gene groups as well as between these groups. Between-group association is suggested by the high $\LL R^2$ values between \textit{FASN1} and both \textit{IM1} and \textit{IM2}. Indeed, Figure \ref{fig:immMet} shows that the expression pattern of \textit{FASN1} is a vertical reflection of these immune response genes' patterns.}}
\label{tab:LLR2}}
\hfill
% -- Table 3 --
\parbox{.48\linewidth}{
\setlength\tabcolsep{4pt}
\begin{tabular}{|c|c|c!{\vrule width 1pt}c|c|c|c|}\hline
\cellcolor{graycell}&  \cellcolor{graycell}{IM1} & \cellcolor{graycell}{IM2} & \cellcolor{graycell}{FASN1} & \cellcolor{graycell}{UGP} & \cellcolor{graycell}{mino} & \cellcolor{graycell}{fbp} \\ \hline
\cellcolor{graycell}{IM1} & - & \cellcolor{redcell}1.00 & \cellcolor{bluecell}0.97 & 0.94 & \cellcolor{bluecell}0.97 & 0.93\\ \hline
\cellcolor{graycell}{IM2} & \cellcolor{redcell}1.00 & - & \cellcolor{bluecell}0.96 & 0.94 & \cellcolor{bluecell}0.97 & 0.93 \\ \noalign{\hrule height 1pt}%
\cellcolor{graycell}{FASN1} & \cellcolor{bluecell}0.97 & \cellcolor{bluecell}0.96 & - & \cellcolor{white}0.91 & \cellcolor{bluecell}0.98 & \cellcolor{white}0.88\\ \hline
\cellcolor{graycell}{UGP} & 0.94 & 0.94 & 0.91 & - & \cellcolor{white}0.93 & \cellcolor{bluecell}0.99 \\ \hline
\cellcolor{graycell}{mino} & \cellcolor{bluecell}0.97 & \cellcolor{bluecell}0.97 & \cellcolor{bluecell}0.98 & 0.93 & - & 0.93 \\ \hline
\cellcolor{graycell}{fbp} & 0.93 & 0.93 & \cellcolor{white}0.88 & \cellcolor{bluecell}0.99 & \cellcolor{white}0.93 & - \\ \hline
\end{tabular}
\caption{\textit{Non-Bayesian $\LL R^2$ values corresponding to each gene pair in Figure \ref{fig:immMet}. Colored cells mark values above 0.96, the 95$^\tth$ percentile of the empirical distribution of this metric for this dataset. Colors have the same interpretation as in Table \ref{tab:LLR2}. Without the proposed Bayesian method, within-group and between-group associations are not identified as clearly as in Table \ref{tab:LLR2}. Furthermore, nearly all values in the table are close to the selected threshold of 0.96, suggesting that $\LL R^2$ values are easily inflated without the use of priors, even for dissimilar temporal expression patterns (such as those of \textit{mino} and \textit{IM1}).\\[2.4em]}}
\label{tab:NonBayesLLR2}}
\end{table}

Within this set of six genes, Table \ref{tab:immMetW} indicates that there is prior evidence of association between the immune response genes \textit{IM1} and \textit{IM2}, as well as between the metabolic genes \textit{mino} and \textit{UGP}. The off-diagonal ``NA'' entries in Table \ref{tab:immMetW} signify that the relationships between the immune and metabolic genes are uncharacterized in the STRING database. However, the Bayesian $\LL R^2$ values between the metabolic gene \textit{FASN1} and both immune response genes are high, as shown in Table \ref{tab:LLR2}, indicating that the relationship identified by \citet{schlamp2020} between these gene groups is automatically uncovered by our proposed Bayesian method. Indeed, Figure \ref{fig:immMet} shows that the temporal expression pattern of \textit{FASN1} resembles a vertically-reflected copy of that of either \textit{IM1} or \textit{IM2}. Table \ref{tab:NonBayesLLR2} shows the non-Bayesian $\LL R^2$ values for each gene pair and demonstrates that computing the $\LL R^2$ without biologically-informed priors yields inflated scores that make it difficult to discern either within- or between-group associations. For further validation of these results, we present Tables \ref{tab:BayesLLR2diff} and \ref{tab:NonBayesLLR2diff} in Appendix \ref{appsec:tables}. These tables display the values of the metric $\LL R^2 - \LL R^2_\own$, which we introduced in Section \ref{sec:clusteringMethod} as a means of assessing whether associations indicated by the $\LL R^2$ alone are false positives. 

\subsection{Biologically-meaningful clustering with the Bayesian $\LL{R^2}$} \label{sec:clusters}

\begin{figure}[H]
\centering\includegraphics[width=\textwidth]{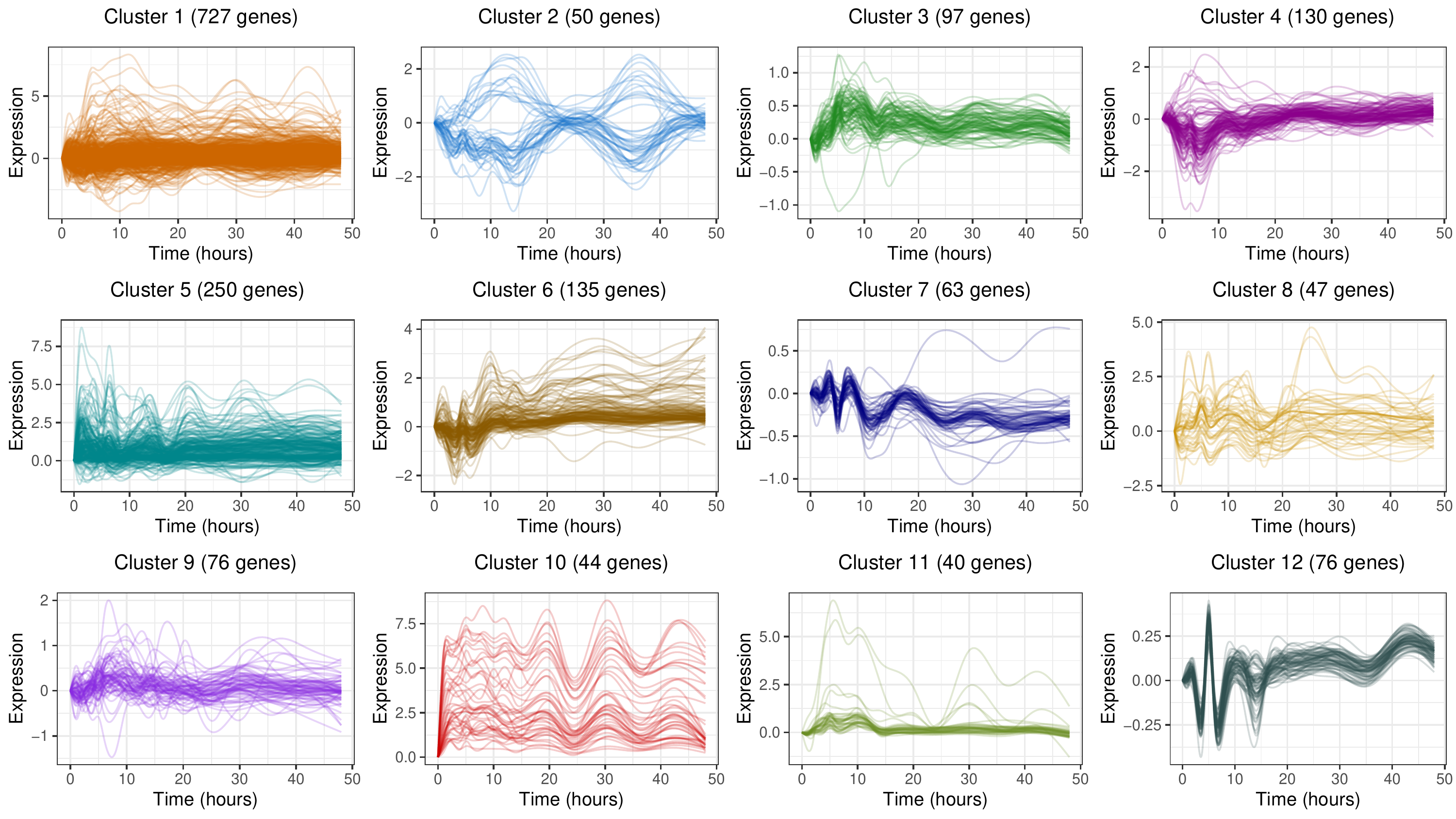}
\caption{\textit{Plots of the temporal expression patterns of genes in each cluster obtained by applying hierarchical clustering to the distance matrix $\mathbf J-\mathbf S$, where $\mathbf S$ contains the Bayesian $\LL R^2$ for all gene pairs. Cluster sizes range from 40 to 727 genes, with a mean of 145 and a median of 76. These plots visually demonstrate that the Bayesian $\LL R^2$ is capable of capturing similarities in the overall shapes of two temporal expression patterns, even if one gene is a time-delayed copy of the other or is reflected across a horizontal axis, for instance.}}
\label{fig:allClusters}
\end{figure}

We now apply hierarchical clustering using Ward's method \citep{ward1963hierarchical} to the $N\times N$ distance matrix $\mathbf J-\mathbf S$, where $\mathbf J$ is a matrix of ones and $\mathbf S_{i,j}=\max\{\LL R^2(i,j), \LL R^2(j,i)\}$ is the similarity matrix containing the symmetrized Bayesian $\LL R^2$ values between all gene pairs. These genes come from the dataset described in Section \ref{sec:dataset}, so $N=1735$. Cutting the resulting dendrogram at a height of ten yields 12 clusters, which we display in Figure \ref{fig:allClusters}. 

From each cluster, we can also construct a network in which an edge is defined between two genes if their corresponding Bayesian $\LL R^2$ exceeds a certain threshold. In our analysis, we choose this threshold to be 0.9, which is also the $99^\tth$ percentile of the empirical distribution of the Bayesian $\LL R^2$ for this dataset.

To understand the biological processes that are represented by these clusters, we make use of the Gene Ontology (GO) resource \citep{ashburner2000gene, carbon2021gene}.  GO links genes with standardized terms that describe their functions. To determine whether a GO term is significantly enriched within a cluster, i.e. whether the cluster contains significantly more genes associated with that term than expected by chance, we perform a hypergeometric test using the R package \texttt{ClusterProfiler} \citep{yu2012clusterprofiler}. We use Benjamini-Hochberg (B-H) corrected $p$-values below 0.05 \citep{benjamini1995controlling} from this test to determine ``significant'' enrichment.

Our analysis shows that with the exception of cluster 8, all clusters are significantly enriched for specific biological functions. Recall that our dataset profiles the dynamics of immune response, which is an energetically costly process that is also associated with metabolic changes \citep{diangelo2009immune}. The interplay between immunity and metabolism, which we briefly explored in Section \ref{sec:immMet}, is represented particularly well in these clusters. Clusters 1, 4, 6, and 7 are significantly enriched for metabolic processes; cluster 10 is significantly enriched for immune response; and cluster 5 is significantly enriched for both metabolic processes and immune responses. Below, we highlight biologically relevant findings from one cluster, and we discuss three additional clusters in Appendix \ref{appsec:results}. These examples show that clustering with the Bayesian $\LL R^2$ allows genes with similar but lagged expression patterns to be grouped together, even in the absence of known prior information. Finally, the Bayesian $\LL R^2$ is not influenced by the direction of gene expression changes (i.e., positive or negative changes), making it easier to detect tradeoffs or negative regulatory interactions between genes. As a comparison, Figures \ref{fig:ClustersNonBayes} and \ref{fig:ClustersCorrelation} in Appendix \ref{appsec:figures} present clustering results obtained using the non-Bayesian $\LL R^2$ as well as Pearson correlation between genes, though neither method groups together genes with similarly-shaped trajectories as effectively. 

\subsubsection{Analysis of cluster 10}

\begin{figure}[H]
\centering\includegraphics[width=10cm]{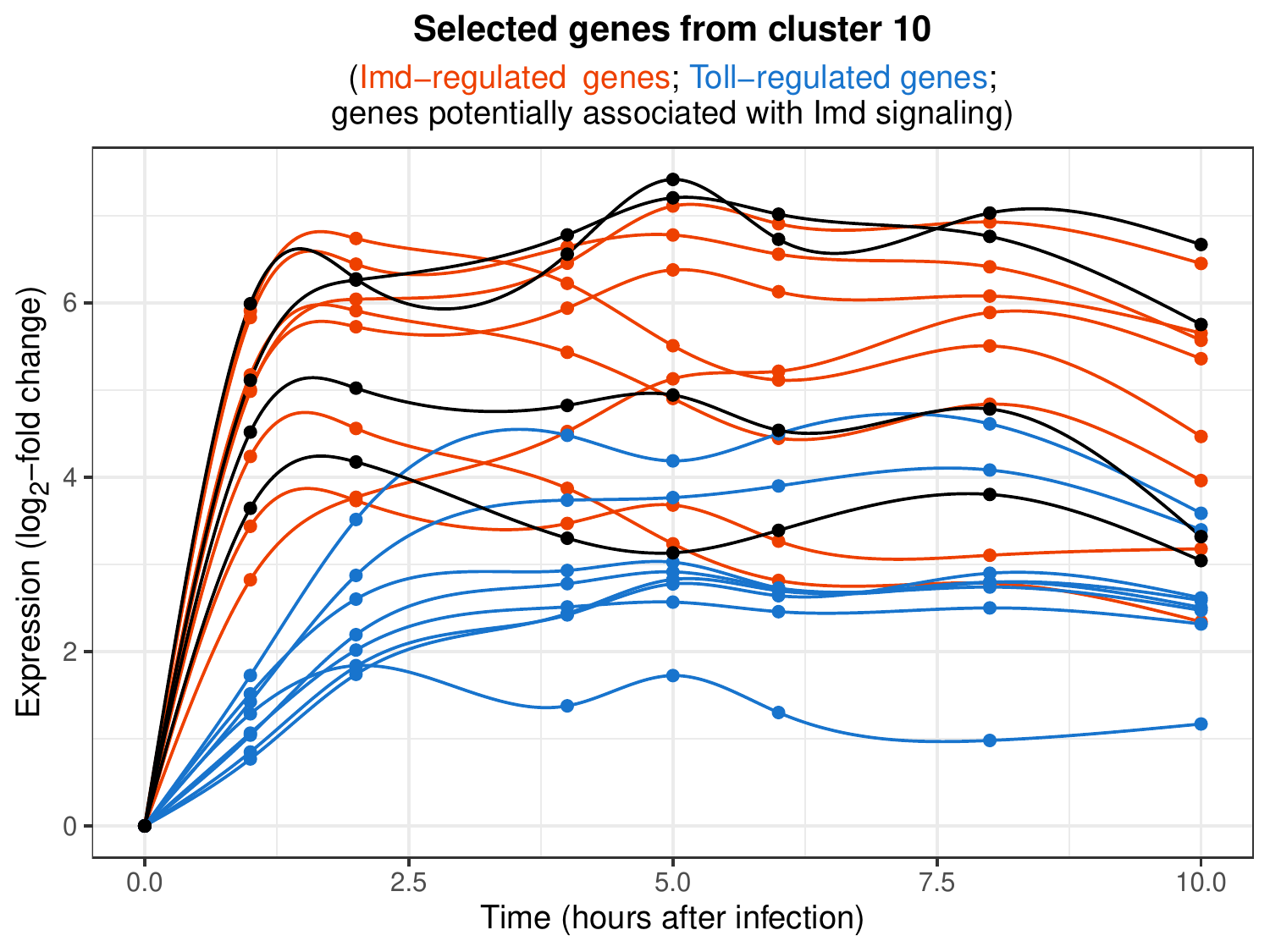}
%\end{figure}
%\begin{figure}[H]
\caption{\textit{Temporal expression patterns of selected genes in cluster 10 during the first ten hours following peptidoglycan injection. The eight red lines correspond to Imd-regulated genes. The eight blue lines correspond to Toll-regulated genes, which show a smaller and delayed up-regulation. The four black lines correspond to genes that exhibited high Bayesian $\LL R^2$ values ($>0.9$) with Imd-regulated genes; while there is no prior information in STRING linking them with the Imd pathway, their co-clustering with Imd-regulated genes was also observed in \citet{schlamp2020}.}}
\label{fig:cluster10}
\end{figure}

In cluster 10, which consists of 44 genes, 
one of the most significantly enriched GO terms is ``defense response''. This GO term is supported by 24 genes in the cluster, within which there are two distinct groups that we display in Figure \ref{fig:cluster10}: eight genes that are known to respond to ``Imd'' signaling and eight genes that respond to ``Toll'' signaling. The Imd and Toll signaling pathways represent well-studied molecular responses to infection in flies. The Imd pathway is tailored to fight off infections from gram-negative bacteria \citep{kaneko2004monomeric, zaidman2011drosophila, hanson2020new}, while the Toll pathway fights off infections from gram-positive bacteria and fungi \citep{gobert2003dual, hanson2020new}.

Since the flies profiled in this gene expression dataset were injected with peptidoglycan derived from \textit{E. coli}, a gram-negative bacterium, we expect to see an activation of Imd-regulated genes. Indeed, as seen in Figure \ref{fig:cluster10}, the eight Imd-regulated genes in this cluster immediately underwent strong up-regulation and reached their highest expression one to two hours after peptidoglycan injection. By contrast, Toll-regulated genes underwent a delayed up-regulation of smaller magnitude, and reached their highest expression two to four hours after injection. Overall, the Bayesian $\LL R^2$ method successfully grouped together functionally related genes with distinct activation kinetics in this cluster. 

In addition to recovering known dynamics of immune response pathways in cluster 10, the Bayesian $\LL R^2$ metric identified several new relationships between genes. In this cluster, four genes (\textit{CG44404}, also known as \textit{IBIN}; \textit{CG43236}, also known as \textit{Mtk-like}; \textit{CG43202}, also known as \textit{BomT1}; and \textit{CG43920}) had no prior information available in the STRING database to link them with Imd-regulated genes. However, these four genes exhibit similar expression patterns to Imd-regulated genes, as seen in Figure \ref{fig:cluster10}, although previous studies examine \textit{CG44404/IBIN} and \textit{CG43202/BomT1} expression downstream of Toll signaling \citep{clemmons2015effector, valanne2019immune}. This suggests that these four genes are not exclusively controlled by Toll signaling, and that they can also respond to Imd signaling after a gram-negative immune challenge. While Imd-regulation of \textit{CG43236/Mtk-like} and \textit{CG43920} has not been experimentally validated, their co-clustering pattern with Imd-regulated genes was also observed by \citet{schlamp2020}. 

\begin{figure}[H]
\centering\includegraphics[width=8cm]{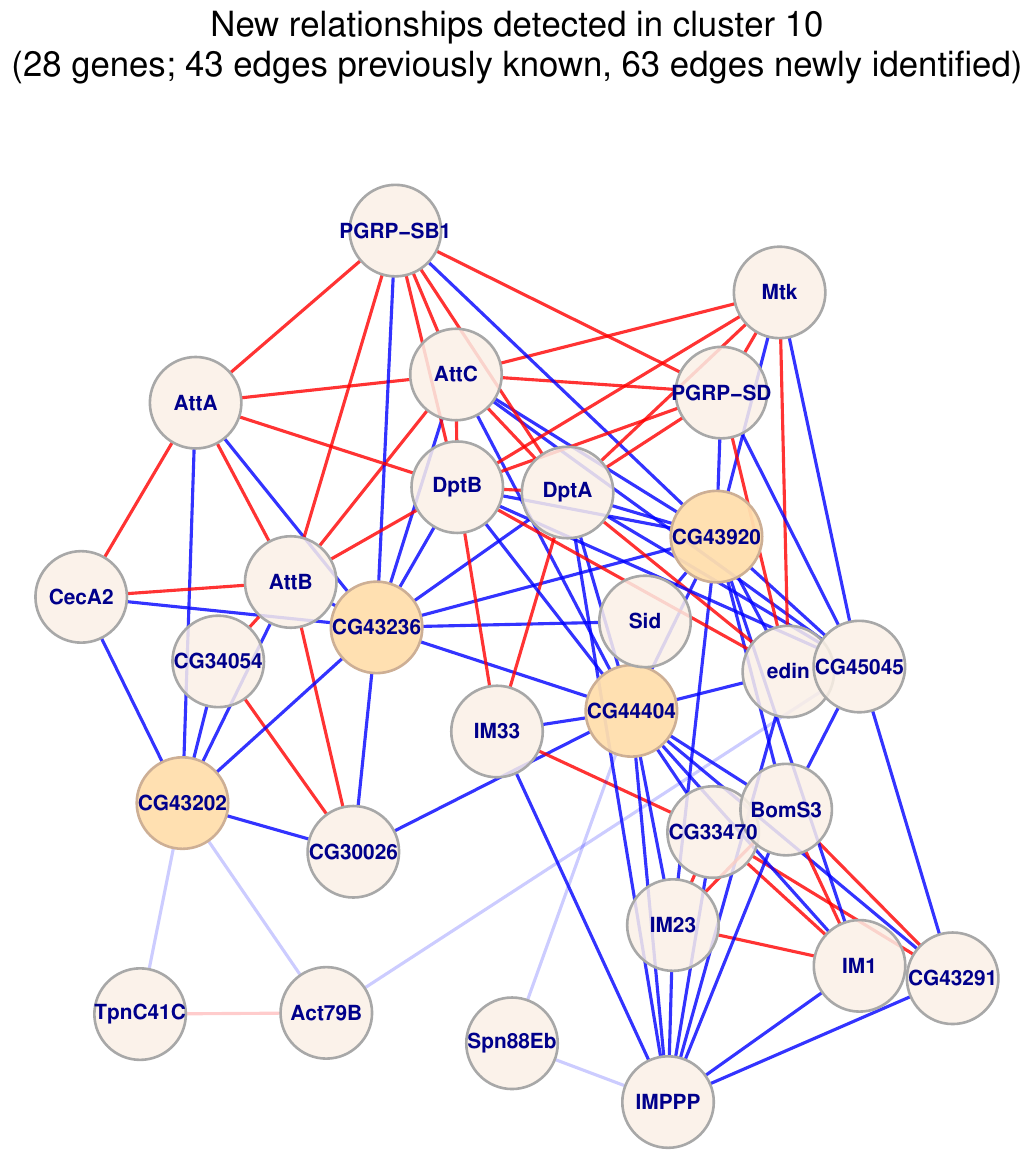}
%\end{figure}
%\begin{figure}[H]
\caption{\textit{Network of genes formed from \textit{CG44404/IBIN, CG43236/Mtk-like, CG43202/BomT1}, \textit{CG43920}, and their neighbors. Two genes are connected by an edge if their Bayesian $\LL R^2 >0.9$. Red and blue edges connect genes with known and unknown relationships, respectively. Darkened edges connect genes within cluster 10. Blue edges connect the four genes of interest with genes known to be regulated by the Imd signaling pathway, suggesting a possible role for them in fighting gram-negative infections.}}
\label{fig:cluster10network}	
\end{figure}

In Figure \ref{fig:cluster10network}, we show a network consisting of the four aforementioned genes (\textit{CG44404/IBIN, CG43236/Mtk-like, CG43202/BomT1, CG43920}) and their neighbors, i.e. the genes with which they have a Bayesian $\LL R^2$ of at least 0.9. Red edges in Figure \ref{fig:cluster10network} connect genes that were known to be associated according to prior information, i.e. $\mathbf W_{ij}=1$ for these pairs. Blue edges connect genes with an uncharacterized relationship in the STRING database, i.e $\mathbf W_{ij}=\text{NA}$ for these pairs.

\subsection{The Bayesian $\LL R^2$ produces a sparse set of associations}\label{sec:R2scatter}

The lead-lag $R^2$ can be computed with biological information via our proposed Bayesian methodology, or without such information via ordinary least-squares regression. We now examine how the Bayesian approach changes the distribution of this quantity in a way that is conducive to identifying pairs or groups of genes with highly similar time dynamics.

In Section \ref{sec:clusteringMethod}, we introduced the metrics $\LL R^2_\other$ and $\LL R^2 - \LL R^2_\own$. As described therein, the former indicates how much variation in gene $A$'s expression can be explained only through the dynamics of another gene $B$. The latter indicates how much additional variation in gene $A$ can be explained by considering gene $B$, on top of considering gene $A$'s own past trajectory. Intuitively, both of these quantities should be large if the two genes are indeed associated in a way that manifests in highly similar temporal dynamics. 
	
\begin{figure}[H]
\includegraphics[width=\textwidth]{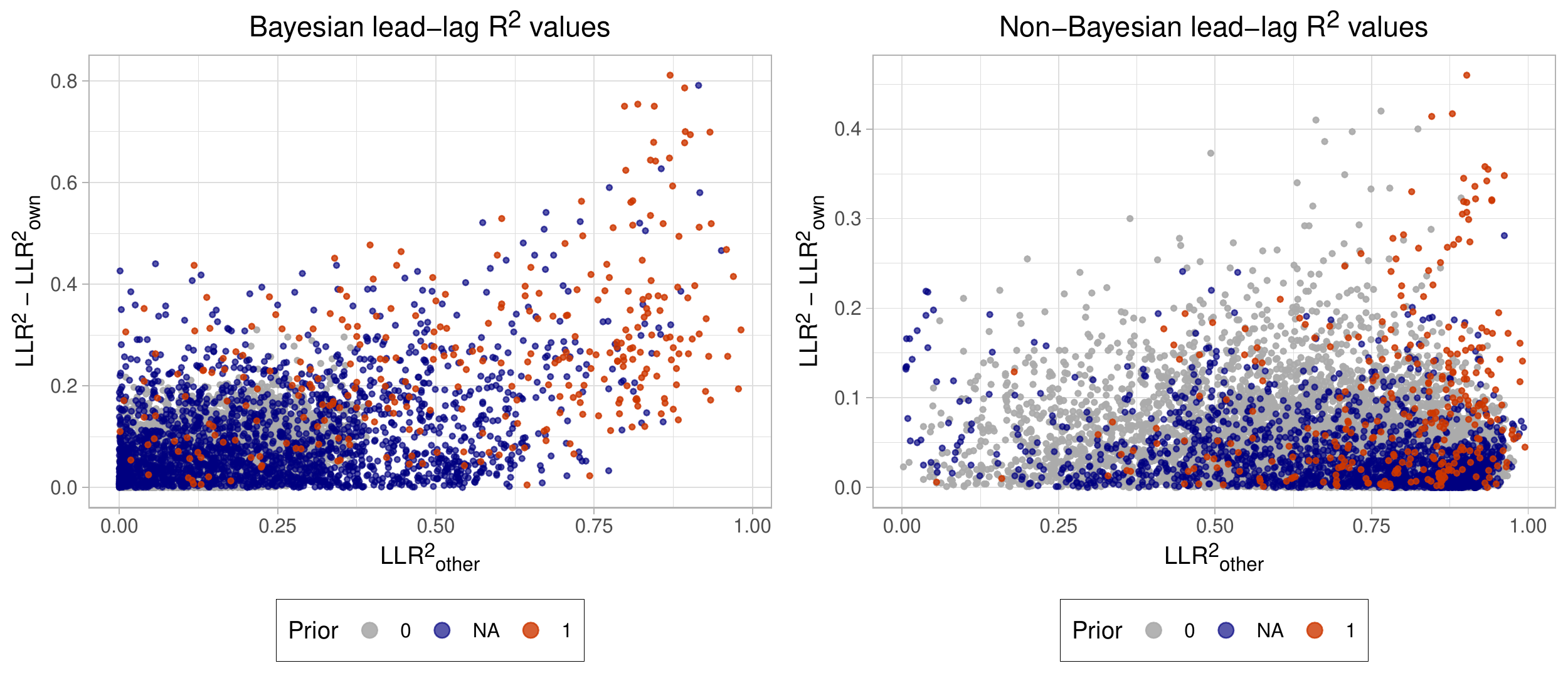}
\caption{\textit{Scatterplots of $\LL R^2_\other$ (horizontal axis) and $\LL R^2-\LL R^2_\own$ (vertical axis) for a random selection of 150 genes, resulting in 11,175 gene pairs. Left: $\LL R^2$ values are computed with the proposed Bayesian approach. Points are colored according to how prior information characterizes the corresponding two genes: unlikely to be associated (gray); uncharacterized association (blue); or known association (red). Right: $\LL R^2$ values are computed via ordinary least-squares, without external biological information.}}
\label{fig:R2scatter}	
\end{figure}

In Figure \ref{fig:R2scatter}, we randomly select 150 genes and place all $\binom{150}{2}=1$1,175 pairs on two scatterplots whose horizontal axes display the $\LL R^2_\other$ values and vertical axes display the $\LL R^2-\LL R^2_\own$ values. These $R^2$ values are computed via our Bayesian method in one scatterplot and via ordinary least-squares regression in the other. Gene pairs of particular interest fall into the upper-right quadrant of the scatterplot.

Figure \ref{fig:R2scatter} shows that when we use the ordinary least-squares approach, i.e. without incorporating external biological information, we obtain an overwhelmingly large number of gene pairs with high $\LL R^2_\other$ scores. Many are those that are unlikely to be associated, according to our chosen sources of prior information. By contrast, the Bayesian approach leverages this information to shift the distribution of the $R^2$ values noticeably, yielding a smaller set of gene pairs that are worth examining further. This distributional shift is due to both the estimator $\bg\beta_*$ in \eqref{eq:betaPostGPrior} for the coefficients in the model (\ref{eq:mA}), as well as the way in which the parameter $g$ is set in (\ref{eq:gCond}). In particular, $g$ controls how much $\bg\beta_*$ is influenced by either the data or prior information.

Importantly, Figure \ref{fig:R2scatter} shows that gene pairs with previously uncharacterized relationships but highly similar time dynamics are more easily identified with the Bayesian $\LL R^2$. A few of these gene pairs, which fall in the fairly sparse upper-right region of the left-hand scatterplot, are shown in Figure \ref{fig:R2scatterMiddle} in Appendix \ref{appsec:figures}. Figure \ref{fig:R2ScatterRight} shows gene pairs in the same region of the scatterplot with well-known relationships, further demonstrating that the proposed method successfully recovers familiar associations.
% TeX root = ../main.pdf

\section{Discussion} \label{sec:conclusion}

Time-course gene expression datasets are a valuable resource for understanding the complex dynamics of interconnected biological systems. An important statistical task in the analysis of such data is to identify clusters or networks of genes that exhibit similar temporal expression patterns. These patterns yield systems-level insight into the roles of biological pathways and processes such as disease progression and recovery. 

The main statistical challenges in studying time-course gene expression datasets stem from their high dimensionality and small sample sizes, combined with the nonlinearity of gene expression time dynamics. To overcome these difficulties, we proposed a method for identifying potentially associated genes that treats temporal gene expression as a dynamical system governed by ODEs, whose parameters are determined in a Bayesian way using gene networks curated \textit{a priori} from biological databases. The ODE model is fit to a pair of genes via Bayesian regression and is used to derive the biologically-informed Bayesian lead-lag $R^2$ similarity measure. The Bayesian regression procedure leverages Zellner's $g$-prior and ideas from shrinkage estimation, namely minimization of Stein's unbiased risk estimate (SURE), to balance the posterior ODE model's fit to the data and to prior information. As a result, we automatically encourage gene pairs with known associations to receive higher Bayesian lead-lag $R^2$ scores, while reducing the scores of gene pairs that are unlikely to be related and allowing new relationships to be discovered.

In Section \ref{sec:results}, we analyzed clusters and networks of genes that were identified by our method as having similar temporal dynamics. In particular, the clusters highlighted the known interplay between immune response and metabolism, and suggested roles for uncharacterized genes displaying remarkably similar temporal patterns to more well-studied ones. We contrasted our results to those obtained by using only the ordinary least-squares version of the lead-lag $R^2$ and demonstrated how the inclusion of prior biological information greatly aids the identification of biologically relevant gene groups.

\clearpage

% --- Appendix ---

\appendix
% TeX root = ../main.tex

\section{Proofs} \label{appsec:proofs}

In this section, we provide proofs of Theorems \ref{thm:SUREoptg} and \ref{thm:sureOptGXi}. 

\begin{proof}[Proof of Theorem \ref{thm:SUREoptg}]
We write $\delta_0(\mathbf Y)$ as $\delta_0(g)$, treating $\mathbf Y$ as fixed. Expanding the expression for $\delta_0(\mathbf g)$, we obtain
\begin{align*}
&\delta_0(g) = \|\mathbf Y - \mathbf X\bg{\hat\beta}\|^2 + \frac{2gp\hat\sigma^2}{1+g}-n\hat\sigma^2 \\
&= \left(\mathbf Y-\frac{1}{1+g}\mathbf Y_0-\frac{g}{1+g}\mathbf{\hat Y}_{\text{OLS}}\right)^T\left(\mathbf Y-\frac{1}{1+g}\mathbf Y_0-\frac{g}{1+g}\mathbf{\hat Y}_{\text{OLS}}\right)+ \frac{2gp\hat\sigma^2}{1+g}-n\hat\sigma^2\\
&= \|\mathbf Y\|^2-\frac{2}{1+g}\mathbf Y^T\mathbf Y_0 - \frac{2g}{1+g}\mathbf Y^T\mathbf{\hat Y}_{\text{OLS}}+\frac{1}{(1+g)^2}\|\mathbf Y_0\|^2 + \frac{2g}{(1+g)^2} \mathbf{\hat Y}_{\text{OLS}}^T \mathbf Y_0 \\
&\hspace{1cm} + \frac{g^2}{(1+g)^2}\|\mathbf{\hat Y}_{\text{OLS}}\|^2 + \frac{2gp\hat\sigma^2}{1+g}-n\hat\sigma^2. \numberthis\label{pf1}
\end{align*}
Next, observe that $\mathbf Y^T\hat{\mathbf Y}_{\text{OLS}} = \|\mathbf{\hat Y}_{\text{OLS}}\|^2$, because
\begin{align}
\|\mathbf{\hat Y}_{\text{OLS}}\|^2 = (\mathbf{HY})^T(\mathbf{HY}) = \mathbf Y^T\mathbf{HY} = \mathbf Y^T\mathbf{\hat Y}_{\text{OLS}}. \label{eq:thm1Id1}
\end{align}
Furthermore, observe that $\mathbf{\hat Y}_{\text{OLS}}^T \mathbf Y_0=\mathbf Y^T\mathbf Y_0$, because
\begin{align}
\mathbf{\hat Y}_{\text{OLS}}^T \mathbf Y_0 = (\mathbf{HY})^T\mathbf Y_0 = \mathbf Y^T\mathbf{HY}_0 = \mathbf Y^T\mathbf Y_0, \label{eq:thm1Id2}
\end{align}
where the last equality follows from the fact that $\mathbf Y_0=\mathbf X\bg\beta_0$, i.e. $\mathbf Y_0$ is in the column span of $\mathbf X$, which is the space onto which $\mathbf H$ projects. The identities (\ref{eq:thm1Id1}) and (\ref{eq:thm1Id2}) can now be used to write $\delta_0(g)$ in (\ref{pf1}) as
\begin{align*}
\delta_0(g) &= \|\mathbf Y\|^2 - \frac{2}{1+g}\mathbf Y^T\mathbf Y_0-\frac{2g}{1+g}\|\mathbf{\hat Y}_{\text{OLS}}\|^2 + \frac{1}{(1+g)^2}\|\mathbf Y_0\|^2 + \frac{2g}{(1+g)^2}\mathbf Y^T\mathbf Y_0 \\
&\hspace{1cm} + \frac{g^2}{(1+g)^2}\|\mathbf{\hat Y}_{\text{OLS}}\|^2 + \frac{2gp\hat\sigma^2}{1+g}-n\hat\sigma^2 \\
&= a-\frac{2b}{1+g} - \frac{2gc}{1+g} + \frac{d}{(1+g)^2}+\frac{2gb}{(1+g)^2}+\frac{g^2c}{(1+g)^2}+\frac{2gp\hat\sigma^2}{1+g}-n\hat\sigma^2,
\end{align*}
where $a=\|\mathbf Y\|^2$, $b=\mathbf Y^T\mathbf Y_0$, $c=\|\mathbf{\hat Y}_{\text{OLS}}\|^2$, and $d=\|\mathbf Y_0\|^2$. Differentiating $\delta_0(g)$ with respect to $g$, we obtain
\begin{align*}
\frac{\dv\delta_0(g)}{\dv g}&= \frac{2b}{(1+g)^2}-\frac{2c}{1+g}+\frac{2gc}{(1+g)^2}-\frac{2d}{(1+g)^3}+\frac{2b}{(1+g)^2}-\frac{4gb}{(1+g)^3} \\
&\hspace{1cm} + \frac{2gc}{(1+g)^2}-\frac{2g^2c}{(1+g)^3}+\frac{2p\hat\sigma^2}{1+g}-\frac{2gp\hat\sigma^2}{(1+g)^2} \\
&= \frac{2p\hat\sigma^2-2c}{1+g} + \frac{4b+4gc-2gp\hat\sigma^2}{(1+g)^2} - \frac{2d+4gb+2g^2c}{(1+g)^3} \\
&= \frac{2p\hat\sigma^2-2c+2gp\hat\sigma^2-2gc}{(1+g)^2} + \frac{4b+4gc-2gp\hat\sigma^2}{(1+g)^2} - \frac{2d+4gb+2g^2c}{(1+g)^3} \\
&= \frac{4b+2gc+2p\hat\sigma^2-2c}{(1+g)^2} - \frac{2d+4gb+2g^2c}{(1+g)^3}.
\end{align*}
Setting this derivative to zero and rearranging yields:
\begin{align*}
\frac{2d+4gb+2g^2c}{(1+g)^3} &\;=\; \frac{4b+2gc+2p\hat\sigma^2-2c}{(1+g)^2} \\
\Rightarrow\;\; 2d+4gb+2g^2c &\;=\; (1+g)\left(4b+2gc+2p\hat\sigma^2-2c\right) \\
&\;=\;4b+2gc+2p\hat\sigma^2-2c+4gb+2g^2c+2gp\hat\sigma^2-2gc \\
\Rightarrow\;\; 2d&\;=\; 4b+2p\hat\sigma^2-2c+2gp\hat\sigma^2\\
\Rightarrow\;\; g_* &\;=\; \frac{c+d-2b}{p\hat\sigma^2} - 1.
\end{align*}
We now substitute the definitions of $b$, $c$, and $d$ back into this expression to obtain
\begin{align}
g_* = \frac{\|\mathbf{\hat Y}_{\text{OLS}}\|^2+\|\mathbf Y_0\|^2-2\mathbf Y^T\mathbf Y_0}{p\hat\sigma^2} - 1. \label{pf2}
\end{align}
The numerator of (\ref{pf2}) can be simplified by observing that
\begin{align*}
\|\mathbf{\hat Y}_{\text{OLS}}\|^2+\|\mathbf Y_0\|^2-2\mathbf Y^T\mathbf Y_0 &= \|\mathbf{\hat Y}_{\text{OLS}}\|^2+\|\mathbf Y_0\|^2-2\mathbf {\hat Y}_{\text{OLS}}^T\mathbf Y_0 \\
&= (\mathbf{\hat Y}_{\text{OLS}}-\mathbf Y_0)^T(\mathbf{\hat Y}_{\text{OLS}}-\mathbf Y_0)^T \\
&= \|\mathbf{\hat Y}_{\text{OLS}}-\mathbf Y_0\|^2.
\end{align*}
Therefore, (\ref{pf2}) becomes
\begin{align}
g_* = \frac{\|\mathbf{\hat Y}_{\text{OLS}}-\mathbf Y_0\|^2}{p\hat\sigma^2} - 1. \label{pf3}
\end{align}
When $\bg\beta_0=\mathbf 0$, we have $\mathbf Y_0 = \mathbf X\bg\beta_0 = \mathbf 0$. In this case, (\ref{pf3}) becomes
\begin{align*}
g_* = \frac{\|\mathbf{\hat Y}_{\text{OLS}}\|^2}{p\hat\sigma^2} - 1.
\end{align*}
Finally, the second derivative of $\delta_0(g)$ evaluated at $g=g_*$ in (\ref{pf3}) is equal to 
$$\frac{\dv^{\:2}\delta_0(g)}{\dv g^2}\bigg|_{g=g_*}=\frac{2p^4\hat\sigma^8}{\|\mathbf{\hat Y}_{\text{OLS}}-\mathbf Y_0\|^6},$$
which is positive, thus confirming that $\delta_0(g)$ is indeed minimized at $g=g_*$. 	This second derivative calculation was verified in \citetalias{Mathematica}.
\end{proof}

\begin{proof}[Proof of Theorem \ref{thm:sureOptGXi}]
We write $\delta_0(\mathbf Y)$ as $\delta_0(g,\xi)$, treating $\mathbf Y$ as fixed. First, if $\bg\beta_0=[\xi,\xi,0,0,0]^T$, then $\mathbf X\bg\beta_0 = \xi\mathbf X_{12}$, where $\mathbf X_{12}$ is the element-wise sum of the first two columns of $\mathbf X$. We now proceed similarly to the proof of Theorem \ref{thm:SUREoptg} by expanding $\delta_0(g,\xi)$:
\begin{align*}
&\delta_0(g,\xi) = \|\mathbf Y-\mathbf X\bg{\hat\beta}\|^2 + \frac{2gp\hat\sigma^2}{1+g}-n\hat\sigma^2 \\
&= \left\|\mathbf Y-\mathbf X\left(\frac{1}{1+g}\bg\beta_0+\frac{g}{1+g}\bg{\hat\beta}_{\text{OLS}}\right)\right\|^2+\frac{2gp\hat\sigma^2}{1+g}-n\hat\sigma^2 \\
&= \left\|\mathbf Y-\frac{\xi}{1+g}\mathbf X_{12}-\frac{g}{1+g}\mathbf{\hat Y}_{\text{OLS}}\right\|^2+\frac{2gp\hat\sigma^2}{1+g}-n\hat\sigma^2 \\
&= \left(\mathbf Y-\frac{\xi}{1+g}\mathbf X_{12}-\frac{g}{1+g}\mathbf{\hat Y}_{\text{OLS}}\right)^T\left(\mathbf Y-\frac{\xi}{1+g}\mathbf X_{12}-\frac{g}{1+g}\mathbf{\hat Y}_{\text{OLS}}\right) + \frac{2gp\hat\sigma^2}{1+g}-n\hat\sigma^2 \\
&= \|\mathbf Y\|^2 - \frac{2\xi}{1+g}\mathbf Y^T\mathbf X_{12}-\frac{2g}{1+g}\mathbf Y^T\mathbf{\hat Y}_{\text{OLS}} + \frac{\xi^2}{(1+g)^2}\|\mathbf X_{12}\|^2 + \frac{2\xi g}{(1+g)^2}\mathbf{\hat Y}_{\text{OLS}}^T\mathbf X_{12} \\
&\hspace{1cm}+ \frac{g^2}{(1+g)^2}\|\hat{\mathbf Y}_{\text{OLS}}\|^2 + \frac{2gp\hat\sigma^2}{1+g}-n\hat\sigma^2. \numberthis\label{eq:pf2}
\end{align*}
Next, observe that $\mathbf{\hat Y}_{\text{OLS}}^T\mathbf X_{12}=\mathbf Y^T\mathbf X_{12}$, because
\begin{align}
\mathbf{\hat Y}_{\text{OLS}}^T\mathbf X_{12} = (\mathbf H\mathbf Y)^T\mathbf X_{12} = \mathbf Y^T\mathbf H\mathbf X_{12} = \mathbf Y^T\mathbf X_{12}, \label{eq:thm2Id1}
\end{align}
where the last equality follows from the fact that $\mathbf X_{12}$ is in the column span of $\mathbf X$, which is the space onto which $\mathbf H$ projects. The identities (\ref{eq:thm1Id1}) and (\ref{eq:thm2Id1}) can now be used to write $\delta_0(g,\xi)$ in (\ref{eq:pf2}) as
\begin{align*}
\delta_0(g,\xi) &= \|\mathbf Y\|^2 - \frac{2\xi}{1+g}\mathbf Y^T\mathbf X_{12} - \frac{2g}{1+g}\|\mathbf{\hat Y}_{\text{OLS}}\|^2 + \frac{\xi^2}{(1+g)^2}\|\mathbf X_{12}\|^2 + \frac{2\xi g}{(1+g)^2}\mathbf Y^T\mathbf X_{12} \\
&\hspace{1cm} + \frac{g^2}{(1+g)^2}\|\mathbf{\hat Y}_{\text{OLS}}\|^2 + \frac{2gp\hat\sigma^2}{1+g}-n\hat\sigma^2 \\
&= a- \frac{2\xi b}{1+g}-\frac{2gc}{1+g}+\frac{\xi^2 d}{(1+g)^2}+\frac{2\xi gb}{(1+g)^2} + \frac{g^2 c}{(1+g)^2} + \frac{2gp\hat\sigma^2}{1+g}-n\hat\sigma^2,
\end{align*}
where $a=\|\mathbf Y\|^2$, $b=\mathbf Y^T\mathbf X_{12}$, $c= \|\mathbf{\hat Y}_{\text{OLS}}\|^2$, and $d=\|\mathbf X_{12}\|^2$. Differentiating $\delta_0(g,\xi)$ with respect to $\xi$, we obtain
\begin{align*}
\frac{\partial\delta_0(g,\xi)}{\partial\xi} &= -\frac{2b}{1+g} + \frac{2\xi d}{(1+g)^2} + \frac{2gb}{(1+g)^2}.
\end{align*}
Setting this derivative to zero and rearranging yields:
\begin{align*}
\frac{2b}{1+g} &\;=\; \frac{2\xi d + 2gb}{(1+g)^2} \\
\Rightarrow\;\; b(1+g) &\;=\; \xi d + gb \\
\Rightarrow\;\; b&\;=\; \xi d \\
\Rightarrow\;\; \xi_* &\;=\; \frac{b}{d}.
\end{align*}
We now substitute the definitions of $b$ and $d$ back into this expression to obtain
\begin{align}
\xi_* = \frac{\mathbf Y^T\mathbf X_{12}}{\|\mathbf X_{12}\|^2}. \label{eq:xiPf}
\end{align}
Next, we differentiate $\delta_0(g,\xi)$ with respect to $g$:
\begin{align*}
\frac{\partial\delta_0(\mathbf Y)}{\partial g} &= \frac{2\xi b}{(1+g)^2} -\frac{2c}{1+g} + \frac{2gc}{(1+g)^2} - \frac{2\xi^2 d}{(1+g)^3} + \frac{2\xi b}{(1+g)^2} - \frac{4\xi gb}{(1+g)^3} \\
&\hspace{1cm}+ \frac{2gc}{(1+g)^2} - \frac{2g^2c}{(1+g)^3} + \frac{2p\hat\sigma^2}{1+g} - \frac{2gp\hat\sigma^2}{(1+g)^2} \\
&= \frac{2p\hat\sigma^2-2c}{1+g} + \frac{4\xi b + 4gc - 2gp\hat\sigma^2}{(1+g)^2} - \frac{2\xi^2 d + 4\xi gb + 2g^2 c}{(1+g)^3} \\
&= \frac{2p\hat\sigma^2-2c + 2gp\hat\sigma^2-2gc}{(1+g)^2} + \frac{4\xi b + 4gc - 2gp\hat\sigma^2}{(1+g)^2} - \frac{2\xi^2 d + 4\xi gb + 2g^2c}{(1+g)^3} \\
&= \frac{4\xi b + 2gc + 2p\hat\sigma^2-2c}{(1+g)^2} - \frac{2\xi^2 d + 4\xi gb+2g^2 c}{(1+g)^3}.
\end{align*}
Setting this derivative to zero and rearranging yields:
\begin{align*}
\frac{2\xi^2 d + 2\xi gb + 2g^2 c}{(1+g)^3} &\;=\; \frac{4\xi b + 2gc + 2p\hat\sigma^2 -2c}{(1+g)^2} \\
\Rightarrow\;\; 2\xi^2 d + 4\xi gb + 2g^2 c &\;=\; (1+g)(4\xi b + 2gc+2p\hat\sigma^2-2c) \\
&\;=\; 2\xi b + 2gc + 2p\hat\sigma^2 - 2c + 4\xi gb + 2g^2 c + 2gp\hat\sigma^2 -2gc\\
\Rightarrow\;\; 2\xi^2 d &\;=\; 4\xi b+2p\hat\sigma^2-2c+2gp\hat\sigma^2 \\
\Rightarrow\;\; g_* &\;=\;\frac{c+\xi^2 d-2\xi b}{p\hat\sigma^2}-1.
\end{align*}
We now substitute the definitions of $b$, $c$, and $d$ back into this expression to obtain
\begin{align*}
g_* = \frac{\|\mathbf{\hat Y}_{\text{OLS}}\|^2 + \xi^2\|\mathbf X_{12}\|^2-2\xi \mathbf Y^T\mathbf X_{12}}{p\hat\sigma^2}-1.
\end{align*}
Substituting $\xi$ in (\ref{eq:xiPf}) into this expression for $g$ yields
\begin{align*}
g_* &= \frac{\|\mathbf{\hat Y}_{\text{OLS}}\|^2 + \left(\frac{\mathbf Y^T\mathbf X_{12}}{\|\mathbf X_{12}\|^2}\right)^2\|\mathbf X_{12}\|^2-2\left(\frac{\mathbf Y^T\mathbf X_{12}}{\|\mathbf X_{12}\|^2}\right)\mathbf Y^T\mathbf X_{12}}{p\hat\sigma^2} -1 \\
&= \frac{\|\mathbf{\hat Y}_{\text{OLS}}\|^2 + \frac{(\mathbf Y^T\mathbf X_{12})^2}{\|\mathbf X_{12}\|^2}-2\frac{(\mathbf Y^T\mathbf X_{12})^2}{\|\mathbf X_{12}\|^2}}{p\hat\sigma^2} -1 \\
&= \frac{\|\mathbf{\hat Y}_{\text{OLS}}\|^2\|\mathbf X_{12}\|^2 - (\mathbf Y^T\mathbf X_{12})^2}{\|\mathbf X_{12}\|^2 p\hat\sigma^2} -1. \numberthis\label{eq:gPf2}
\end{align*}
Finally, the determinant of the Hessian matrix of $\delta_0(g,\xi)$, denoted $\nabla^2\delta_0$, evaluated at $g=g_*$ in (\ref{eq:gPf2}) and $\xi=\xi_*$ in (\ref{eq:xiPf}) is
\begin{align*}
\det \nabla^2\delta_0|_{(g_*,\xi_*)} &= \frac{\partial^2\delta_0}{\partial g^2}\bigg|_{(g_*,\xi_*)}\frac{\partial^2\delta_0}{\partial \xi^2}\bigg|_{(g_*,\xi_*)}-\left(\frac{\partial^2\delta_0}{\partial g\partial\xi}\bigg|_{(g_*,\xi_*)}\right)^2 \\[.6em]
&= \frac{-4\|\mathbf X_{12}\|^{12}\:p^6\:\hat\sigma^{12}}{\left((\mathbf Y^T\mathbf X_{12})^2-\|\mathbf{\hat Y}_{\text{OLS}}\|^2\|\mathbf X_{12}\|^2\right)^5}. \numberthis\label{eq:sureHess}
\end{align*}
We must now verify that $\det \nabla^2\delta_0|_{(g_*,\xi_*)} >0$, i.e. that $(g_*,\xi_*)$ is indeed an extremum of $\delta_0$. First, we observe that
\begin{align}
\|\mathbf{\hat Y}_{\text{OLS}}\|^2\|\mathbf X_{12}\|^2 > (\mathbf Y^T\mathbf X_{12})^2, \label{eq:hessCS}
\end{align}
by direct application of the Cauchy-Schwarz inequality (recall that $\mathbf Y^T\mathbf X_{12}=\mathbf{\hat Y}_{\text{OLS}}^T\mathbf X_{12}$ from (\ref{eq:thm2Id1})). The inequality is strict because of the assumption that the entries of $\bg{\hat\beta}_{\text{OLS}}$ are distinct, which prevents $\mathbf{\hat Y}_{\text{OLS}}$ and $\mathbf X_{12}$ from being linearly dependent. Thus, (\ref{eq:hessCS}) implies that the denominator in (\ref{eq:sureHess}) is strictly negative. The numerator is also strictly negative, because the assumption that gene $B$ is not zero everywhere ensures that $\|\mathbf X_{12}\|^2\neq 0$ (recall that $\mathbf X$ is defined in (\ref{eq:responseDesign}), and its first two columns consist of gene $B$'s expression measurements over time and its time integrals). Therefore, (\ref{eq:sureHess}) is strictly positive. To verify that $(g_*,\xi_*)$ is indeed a minimizer of $\delta_0$, we now check that $\frac{\partial^2\delta_0}{\partial\xi^2}|_{(g_*,\xi_*)}>0$ as well. We have:
\begin{align*}
\frac{\partial^2\delta_0}{\partial\xi^2}\bigg|_{(g_*,\xi_*)} = \frac{2\|\mathbf X_{12}\|^6p^2\hat\sigma^4}{\left((\mathbf Y^T\mathbf X_{12})^2-\|\mathbf{\hat Y}_{\text{OLS}}\|^2\|\mathbf X_{12}\|^2\right)^2}
\end{align*}
which is strictly positive. This second derivative calculation, as well as those in (\ref{eq:sureHess}), were verified in \citetalias{Mathematica}.
\end{proof}

\clearpage

% TeX root = ../main.tex

\section{Bayesian lead-lag $R^2$ algorithm} \label{appsec:algo}

Following is an algorithm for computing the Bayesian lead-lag $R^2$ for all gene pairs in a time-course gene expression dataset, consisting of $N$ genes measured at $n$ time points $t_1,...,t_n$.\medskip

{\renewcommand{\baselinestretch}{1}\small\begin{algorithm}[H]
\SetAlgoLined
\KwIn{$\mathbf Z$, a gene expression dataset of size $N\times n$ ($N$ genes observed at $n$ time points); $\mathbf W$, a prior biological information matrix of size $N\times N$; $\mathbf t$, a vector of $n$ time points $t_1,..., t_n$.}
\KwOut{$\mathbf S$, a Bayesian lead-lag $R^2$ similarity matrix of size $N\times N$ (upper-triangular).}
Initialize $\mathbf S$\;
\For{{\normalfont gene} $i = 1,...,N-1$}{

 \For{{\normalfont gene} $j = 2,...,N$}{
\tcp{Get expression measurements for genes $i$ and $j$}
Let $\mathbf x_i^T = \mathbf Z[i,:]$ and $\mathbf x_j^T=\mathbf Z[j,:]$\;\vspace{.5em}

\tcp{Get spline interpolants of given expression measurements}
Let $s_i(t)$ = \texttt{spline}($\mathbf x_i$) and $s_j(t)$ = \texttt{spline}($\mathbf x_j$)\; \vspace{0.5em}

\tcp{Numerically integrate spline interpolants up to each time point}
Let $\tilde{\mathbf s}_i = [\int_0^{t_1} s_i(t)\dv t,\;...,\; \int_0^{t_n} s_i(t)\dv t]^T$ and $\tilde{\mathbf s}_j = [\int_0^{t_1} s_j(t)\dv t,\;...,\; \int_0^{t_n} s_j(t)\dv t]^T$\; \vspace{0.5em}

\tcp{Define $n\times 5$ matrix $\mathbf X$ and $n\times 1$ vector $\mathbf Y$ according to (\ref{eq:responseDesign})}
Let $\mathbf X = [\mathbf x_j \ \ \ \tilde{\mathbf s}_j \ \ \  \tilde{\mathbf s}_i \ \ \ \mathbf t \ \ \ \1_n ]$ and $\mathbf Y = \mathbf x_i$\;\vspace{0.5em}

\tcp{Define prior mean of regression coefficients}
\eIf{$\mathbf W_{ij}=1$}{
Let $\bg\beta_0 = [1,1,0,0,0]^T$\;
}
{
Let $\bg\beta_0 = [0,0,0,0,0]^T$\;
}

\vspace{0.5em}\tcp{Compute least-squares estimates}
Let $\bg{\hat \beta}_{\text{OLS}} = (\mathbf X^T\mathbf X)^{-1}\mathbf X^T\mathbf Y$\; \vspace{0.5em}

\tcp{Compute $g$ in Zellner's $g$-prior via Theorem \ref{thm:SUREoptg} and (\ref{eq:gCond})}
Let $\mathbf Y_0 = \mathbf X\bg\beta_0$ and $\mathbf{\hat Y}_\text{OLS} = \mathbf X\bg{\hat\beta}_\text{OLS}$\;
Let $\hat\sigma^2 = \|\mathbf Y-\mathbf{\hat Y}_\text{OLS}\|^2/(n-p)$\;
\eIf{$\mathbf W_{ij}=1$ \textbf{or} $\mathbf W_{ij}=\text{NA}$}{
Let $g = (\|\mathbf{\hat Y}_\text{OLS}-\mathbf Y_0\|^2/p\hat\sigma^2)-1$\;
}
{
Let $g=1$\;
}

\tcp{Compute posterior mean of regression coefficients according to (\ref{eq:betaPostGPrior})}
Let $\bg\beta_* = (1/(1+g))\bg\beta_0 + (g/(1+g))\bg{\hat\beta}_\text{OLS}$\;\vspace{0.5em}

\tcp{Compute Bayesian lead-lag $R^2$}
Let $\mathbf S_{ij} = \widehat\var(\mathbf X\bg\beta_*) \:/ (\widehat\var(\mathbf X\bg\beta_*) + \widehat\var(\mathbf Y - \mathbf X\bg\beta_*))$\;
}
}
\caption{Bayesian lead-lag $R^2$ calculations for all gene pairs}
\end{algorithm}}

\medskip Note that this algorithm produces an upper-triangular similarity matrix $\mathbf S$ resulting from regressing gene $i$ on gene $j$, for all $i<j$, and storing the resulting Bayesian $\LL R^2(i,j)$ value. In the empirical analysis presented in this paper, we set $\mathbf S_{i,j}=\max\{\LL R^2(i,j), \LL R^2(j,i)\}$. 

To instead compute the Bayesian $\LL R^2_\other$ from sub-model 1 in (\ref{eq:subModel1}), it suffices to change the definition of $\mathbf X$ to $[\mathbf x_j\ \  \tilde{\mathbf s}_j\ \ \1_n]$, and to define $\bg\beta_0$ as $[1,1,0]^T$ if $\mathbf W_{ij}=1$ and $[0,0,0]^T$ otherwise. To compute the Bayesian $\LL R^2_\own$ from sub-model 2 in (\ref{eq:subModel2}), we change $\mathbf X$ to $[\tilde{\mathbf s}_i \ \ \mathbf t \ \ \1_n]$, and $\bg\beta_0=[0,0,0]$ regardless of the value of $\mathbf W_{ij}$. An optimized version of this algorithm runs in 21.8 minutes for $N=1735$ genes on a 2017 3.1 GHz Intel Core i5 MacBook Pro.

% TeX root = ../main.tex

\section{Additional dataset details} \label{appsec:data}

In this section, we provide further details on how our primary time-course gene expression dataset was constructed.

We first reduce our set of 12,657 genes into a set of 951 ``differentially-expressed'' (DE) genes identified in \citet{schlamp2020}, defined as genes satisfying either of the following criteria: 1) there is at least one time point at which the gene's expression undergoes a $\log_2$-fold change of at least two, or 2) a spline-based method for differential expression analysis returns a significant result. The latter method involves fitting a cubic spline to the temporal expression measurements of a gene under treatment and control settings, and testing whether the difference in the resulting two sets of coefficients is significant. We then add back a set of 784 genes that are not DE by these criteria, but that are ``neighbors'' of at least one DE gene. We define a neighbor of a DE gene as a non-DE gene that has a STRING score of at least 0.95 with the DE gene. The purpose of adding such neighbors back into the dataset, now consisting  of $951 + 784 =1735$ genes, is to enable more complete biological pathways to be reconstructed from our cluster analysis. 

In Section \ref{sec:bioPriors}, we describe several sources of biological information that can be encoded into a prior adjacency matrix $\mathbf W$. We choose to use the STRING database, and we mark two genes as ``associated'' (i.e., $\mathbf W_{ij}=1)$ if their STRING score is greater than 0.5. We additionally use replicate information from the time-course dataset to fill in some of the unknown STRING scores. Specifically, if two genes have entries in the STRING database but have an unknown STRING score, we set $\mathbf W_{ij}=1$ if the correlation between their replicated temporal expressions is greater than 0.8. We keep $\mathbf W_{ij}=\text{NA}$ for the gene pairs that do not have entries in STRING.

\section{Additional figures} \label{appsec:figures}

\begin{figure}[H]
\includegraphics[width=\textwidth]{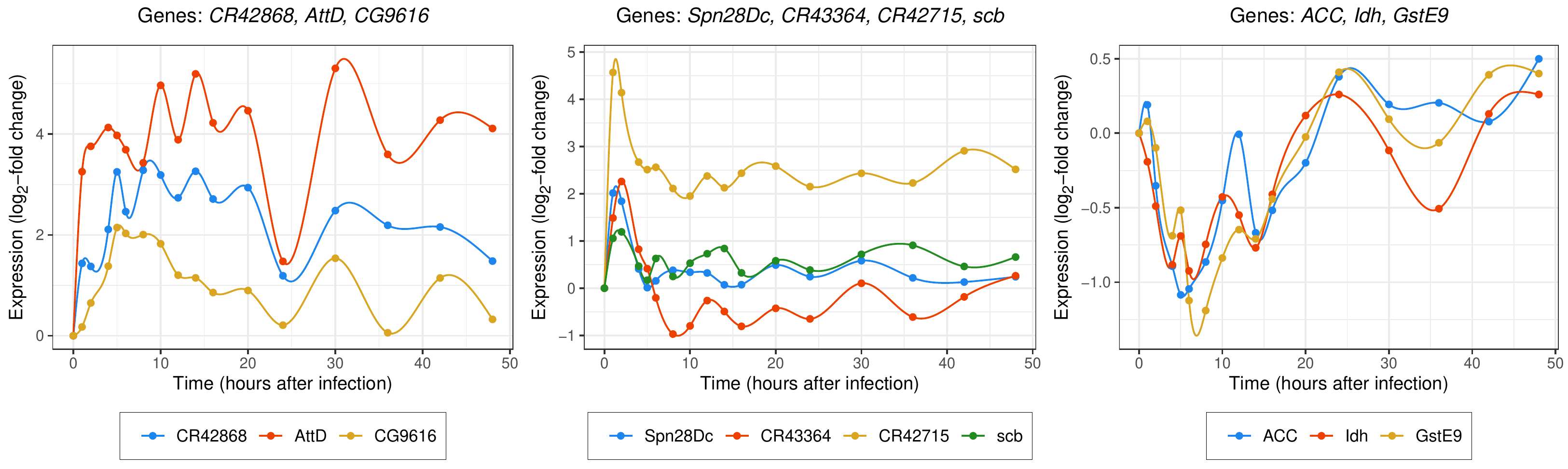}
\end{figure}
\begin{figure}[H]
\caption{\textit{Sets of genes in the upper-right region of the Bayesian $\LL R^2$ scatterplot in Figure \ref{fig:R2scatter}, several of which have uncharacterized pairwise associations. Left: AttD is involved in immune response against Gram-negative bacteria; it exhibits similar patterns to CR42868 and CG9616, neither of which have known molecular functions according to FlyBase \citep{larkin2020flybase}. Middle: Spn28Dc is involved in response to stimuli and protein metabolism, and scb is involved in cell death and organ development. These two genes display similar expression patterns that are also nearly identical in shape to those of the less well-understood RNAs CR43364 and CR42715. Right: Genes ACC, Idh, and GstE9 are involved in a variety of metabolic processes, but not all of their pairwise interactions are known in STRING.}}
\label{fig:R2scatterMiddle}	
\end{figure}

\begin{figure}[H]
\includegraphics[width=\textwidth]{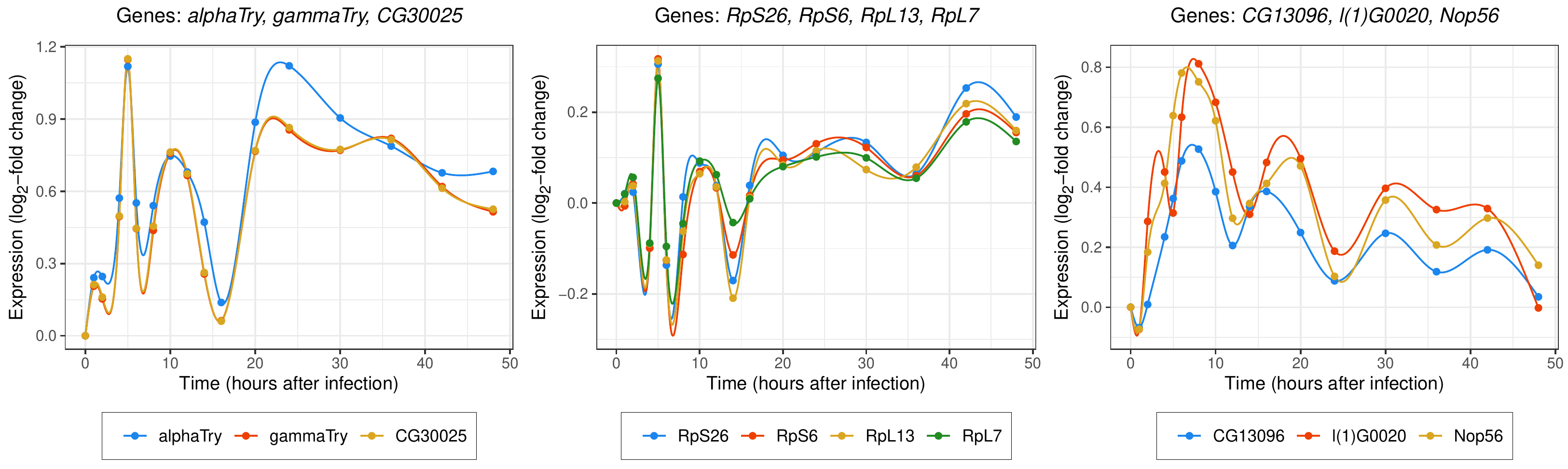}
\caption{\textit{Sets of genes appearing in the upper-right region of the Bayesian $\LL R^2$ scatterplot in Figure \ref{fig:R2scatter}, all of which have known pairwise associations. Left: Genes known to be involved in proteolysis. Middle: Genes that encode ribosomal proteins. Right: Genes known to be involved in RNA binding.}}
\label{fig:R2ScatterRight}	
\end{figure}

\begin{figure}[H]
\includegraphics[width=\textwidth]{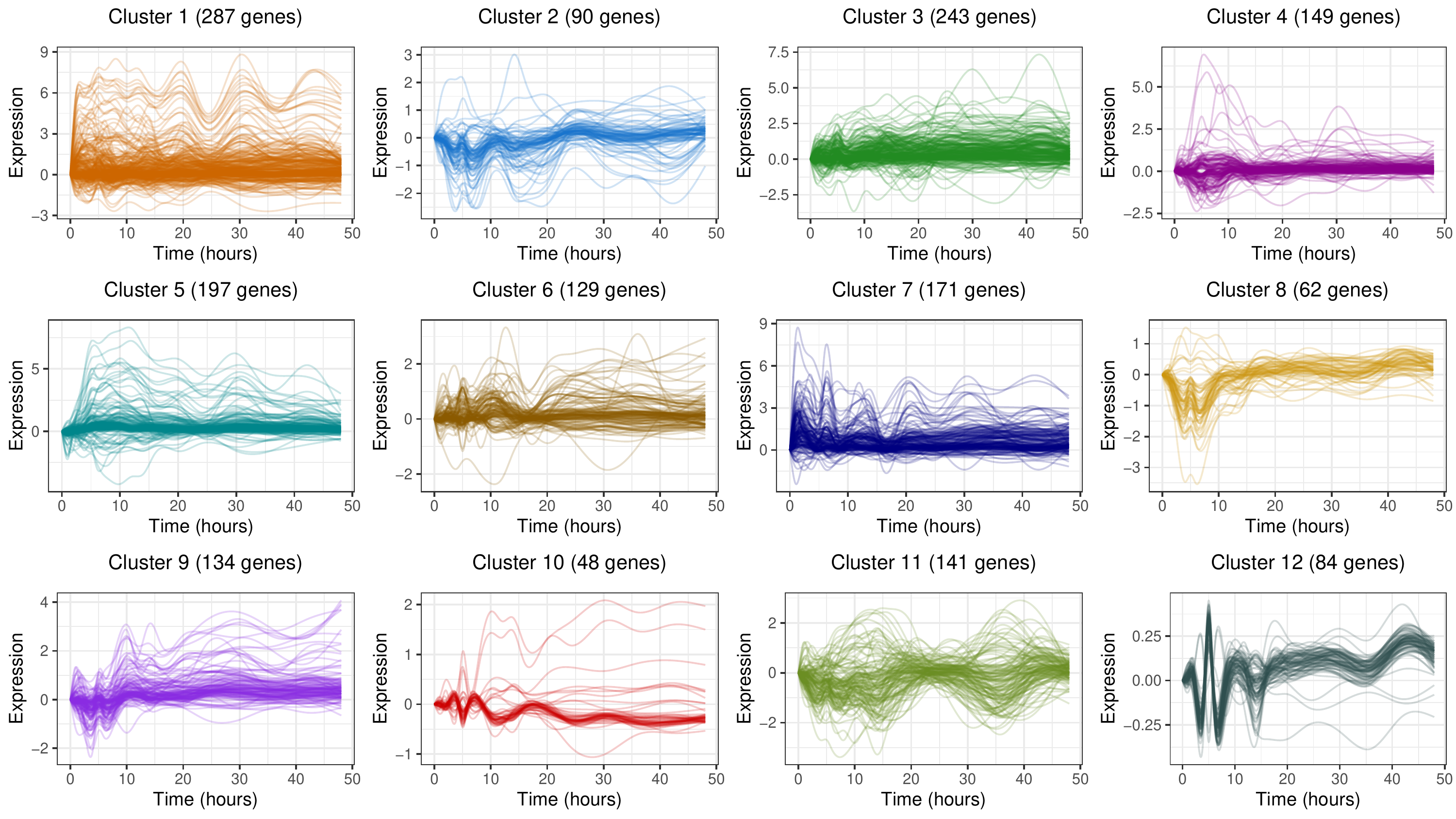}
%\end{figure}
%\begin{figure}[H]
\caption{\textit{Clusters of genes obtained using the non-Bayesian $\LL R^2$ as the similarity metric between genes. Each sub-plot shows the temporal expressions of genes in the corresponding cluster. Clusters were obtained via hierarchical clustering with Ward's method. The dominant pattern in many clusters is either flat or not immediately discernible, indicating that the non-Bayesian $\LL R^2$ is alone insufficient for identifying groups of genes whose temporal patterns are of similar shapes over time.}}
\label{fig:ClustersNonBayes}
\end{figure}

\begin{figure}[H]
\includegraphics[width=\textwidth]{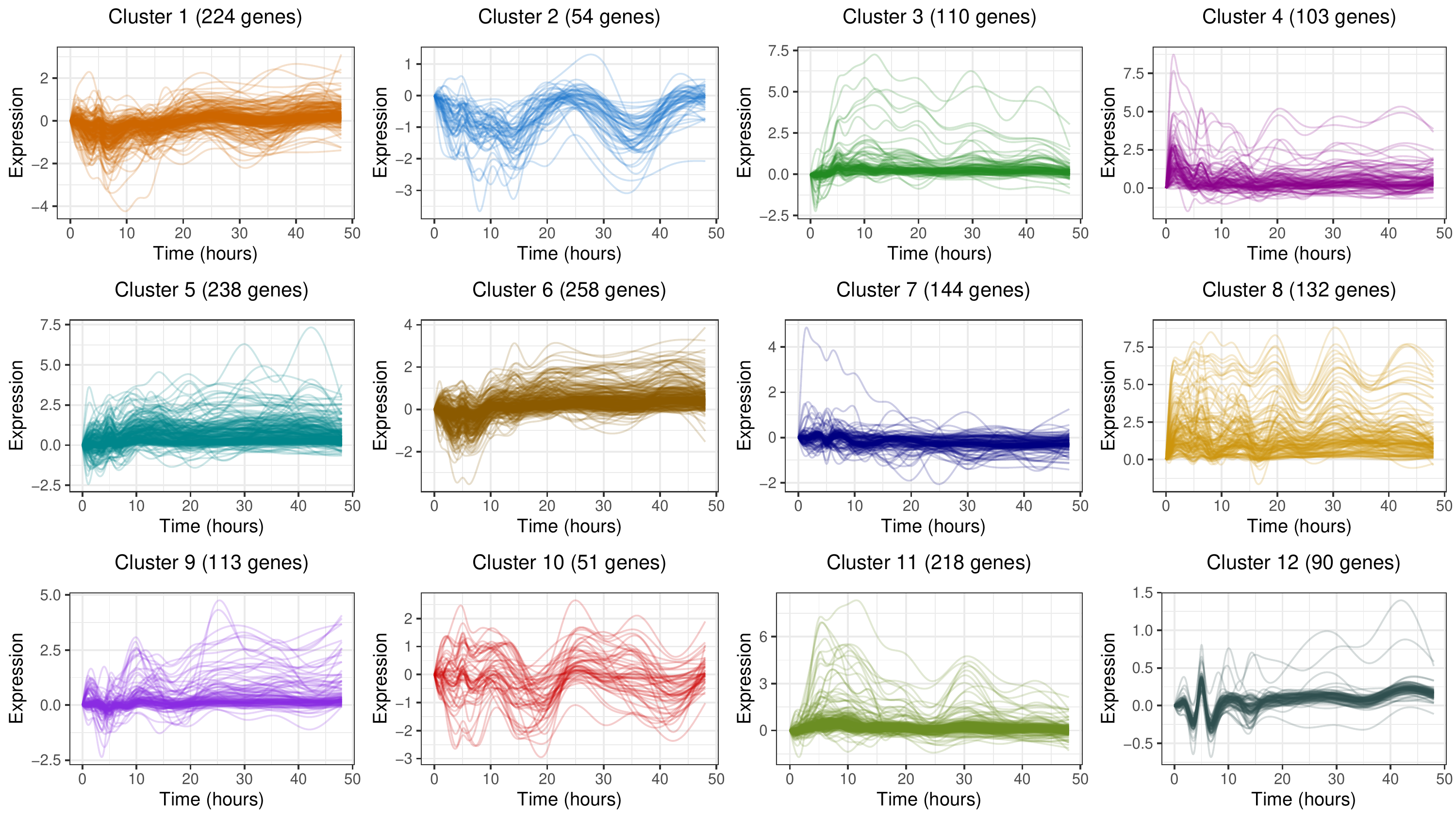}
\caption{\textit{Clusters of genes obtained using Pearson correlation as the similarity metric between genes. Each sub-plot shows the temporal expressions of genes in the corresponding cluster. In cluster and network analyses of time-course gene expression data, Pearson correlation is one of the more commonly used metrics of association. Clusters were obtained via hierarchical clustering with Ward's method. The dominant pattern in many clusters is either flat or not immediately discernible, indicating that correlation is alone insufficient for identifying groups of genes whose temporal patterns are of similar shapes over time.}}
\label{fig:ClustersCorrelation}
\end{figure}

\section{Additional tables} \label{appsec:tables}

\begin{table}[H]
\small\centering
% -- Table 2 --
\parbox{.48\linewidth}{
\setlength\tabcolsep{4pt}
\begin{tabular}{|c|c|c!{\vrule width 1pt}c|c|c|c|}\hline
\cellcolor{graycell}&  \cellcolor{graycell}{IM1} & \cellcolor{graycell}{IM2} & \cellcolor{graycell}{FASN1} & \cellcolor{graycell}{UGP} & \cellcolor{graycell}{mino} & \cellcolor{graycell}{fbp} \\ \hline
\cellcolor{graycell}{IM1} & - & \cellcolor{redcell}0.64 & \cellcolor{bluecell}0.27 & 0.05 & \cellcolor{white}0.09 & 0.09\\ \hline
\cellcolor{graycell}{IM2} & \cellcolor{redcell}0.64 & - & \cellcolor{bluecell}0.26 & 0.04 & \cellcolor{white}0.09 & 0.06 \\ \noalign{\hrule height 1pt}%
\cellcolor{graycell}{FASN1} & \cellcolor{bluecell}0.27 & \cellcolor{bluecell}0.26 & - & \cellcolor{bluecell}0.20 & \cellcolor{bluecell}0.27 & \cellcolor{bluecell}0.23\\ \hline
\cellcolor{graycell}{UGP} & 0.05 & 0.04 & \cellcolor{bluecell}0.20 & - & \cellcolor{redcell}0.20 & \cellcolor{bluecell}0.39 \\ \hline
\cellcolor{graycell}{mino} & 0.09 & 0.09 & \cellcolor{bluecell}0.27 & \cellcolor{redcell}0.20 & - & 0.19 \\ \hline
\cellcolor{graycell}{fbp} & 0.09 & 0.06 & \cellcolor{bluecell}0.23 & \cellcolor{bluecell}0.39 & \cellcolor{white}0.19 & - \\ \hline
\end{tabular}\caption{\textit{Values of the Bayesian $\LL R^2 - \LL R^2_\own$ corresponding to each gene pair in Figure \ref{fig:immMet}. Colored cells mark values above 0.20, the 95$^\tth$ percentile of the empirical distribution of this metric for this dataset. Red cells are consistent with prior evidence of association, and blue cells point to potential associations that may not have been known previously. Most of the colored cells in Table \ref{tab:LLR2} are colored here as well, indicating that the associations drawn from Table \ref{tab:LLR2} are unlikely to be false positives.\\}}
\label{tab:BayesLLR2diff}}
\hfill
% -- Table 3 --
\parbox{.48\linewidth}{
\setlength\tabcolsep{4pt}
\begin{tabular}{|c|c|c!{\vrule width 1pt}c|c|c|c|}\hline
\cellcolor{graycell}&  \cellcolor{graycell}{IM1} & \cellcolor{graycell}{IM2} & \cellcolor{graycell}{FASN1} & \cellcolor{graycell}{UGP} & \cellcolor{graycell}{mino} & \cellcolor{graycell}{fbp} \\ \hline
\cellcolor{graycell}{IM1} & - & \cellcolor{white}0.09 & \cellcolor{white}0.06 & 0.03 & \cellcolor{white}0.06 & 0.02\\ \hline
\cellcolor{graycell}{IM2} & \cellcolor{white}0.09 & - & \cellcolor{white}0.06 & 0.03 & \cellcolor{white}0.06 & 0.02 \\ \noalign{\hrule height 1pt}%
\cellcolor{graycell}{FASN1} & \cellcolor{white}0.06 & \cellcolor{white}0.06 & - & \cellcolor{white}0.09 & \cellcolor{bluecell}0.17 & \cellcolor{bluecell}0.13\\ \hline
\cellcolor{graycell}{UGP} & 0.03 & 0.03 & 0.09 & - & \cellcolor{white}0.11 & \cellcolor{bluecell}0.17 \\ \hline
\cellcolor{graycell}{mino} & 0.06 & 0.06 & \cellcolor{bluecell}0.17 & 0.11 & - & 0.11 \\ \hline
\cellcolor{graycell}{fbp} & 0.02 & 0.02 & \cellcolor{bluecell}0.13 & \cellcolor{bluecell}0.17 & \cellcolor{white}0.11 & - \\ \hline
\end{tabular}
\caption{\textit{Values of the non-Bayesian $\LL R^2 - \LL R^2_\own$ corresponding to each gene pair in Figure \ref{fig:immMet}. Colored cells mark values 0.13, the 95$^\tth$ percentile of the empirical distribution of this metric for this dataset. Colors have the same interpretation as in Table \ref{tab:BayesLLR2diff}. Most of the colored cells in Table \ref{tab:NonBayesLLR2}, which displays the non-Bayesian $\LL R^2$ values, are not highlighted in this table; thus, the non-Bayesian $\LL R^2 - \LL R^2_\own$ metric is less helpful than its Bayesian counterpart in verifying whether or not the associations suggested by the $\LL R^2$ alone are false positives.}}
\label{tab:NonBayesLLR2diff}}
\end{table}

\section{Additional results} \label{appsec:results}

We continue our analysis in Section \ref{sec:clusters} of the results of hierarchical clustering with the Bayesian lead-lag $R^2$ ($\LL R^2$).

\subsection{Analysis of cluster 2}

\begin{figure}[H]
\centering\includegraphics[width=10cm]{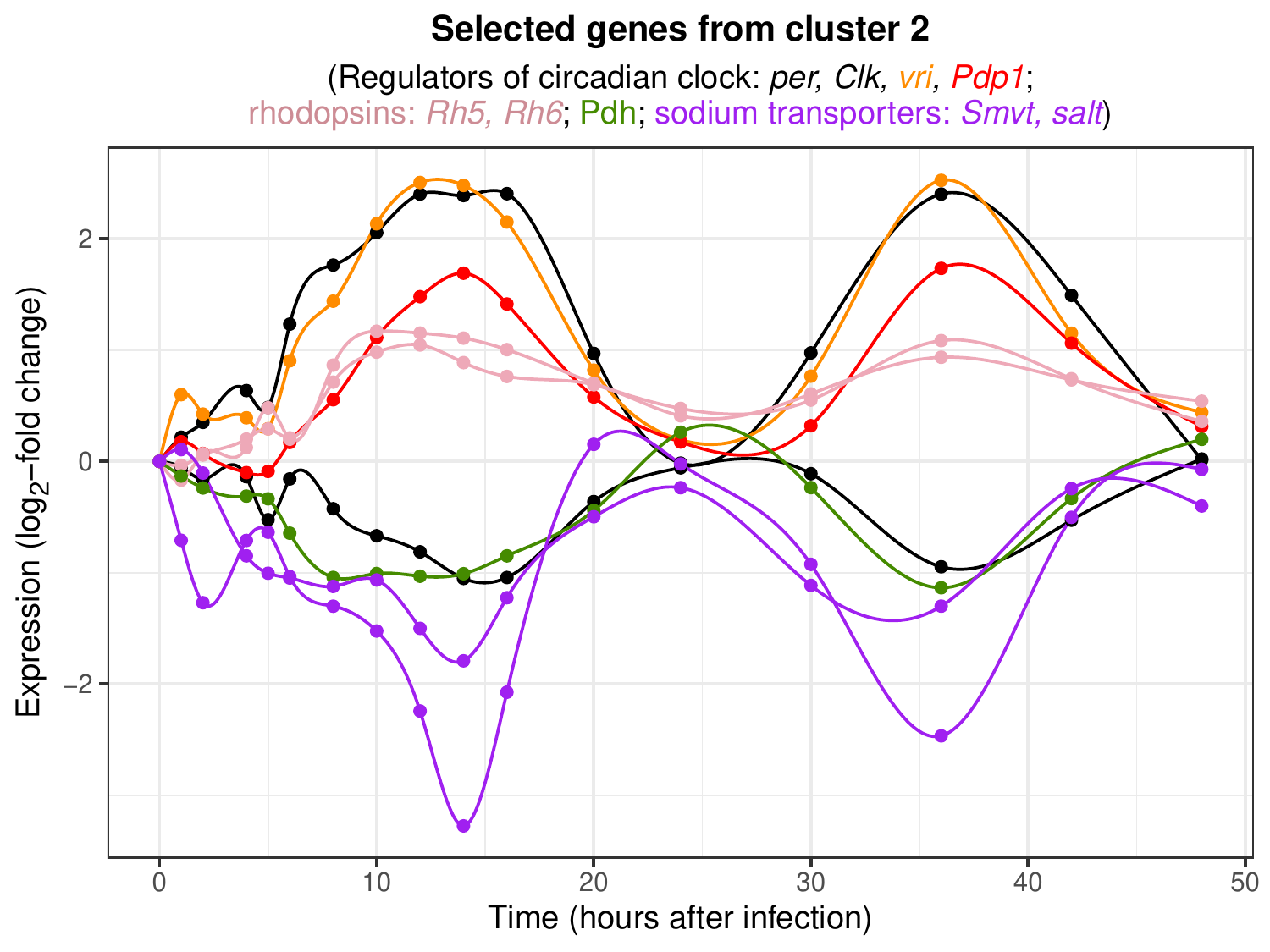}
%\end{figure}
%\begin{figure}[H]
\caption{\textit{Temporal expression patterns of selected genes in cluster 2. Black, red, and orange lines correspond to regulators of the circadian clock (black: \textit{per, Clk}; orange: \textit{vri}; red: \textit{Pdp1}). \textit{Pdp1} is known to reach its peak expression after \textit{vri} does. Pink and green lines correspond to genes that are involved in visual perception (pink: rhodopsins \textit{Rh5, Rh6}; green: retinal pigment dehydrogenase \textit{Pdh}). Purple lines correspond to genes that encode sodium transporters (\textit{Smvt, salt}).}}
\label{fig:cluster2}	
\end{figure}

Cluster 2 contains both up- and down-regulated genes with circadian rhythms, according to GO term enrichment. Several of these genes are displayed in Figure \ref{fig:cluster2}. Among the up-regulated genes are three regulators of the circadian clock (\textit{per, vri, Pdp1}). A fourth regulator of the circadian clock, \textit{Clk}, is down-regulated. \textit{Pdp1} has been reported to reach its peak expression three to six hours after \textit{vri}'s peak expression \citep{cyran2003vrille}, a pattern that is visible in this cluster. Cluster 2 further contains genes that are involved in visual perception: two genes encoding rhodopsins (\textit{Rh5, Rh6}) \citep{gaudet2011phylogenetic} and \textit{Pdh}, which encodes a retinal pigment dehydrogenase \citep{wang2010requirement}. Similar to \citet{schlamp2020}, we also found that \textit{Smvt} and \textit{salt}, which encode sodium transporters \citep{gaudet2011phylogenetic}, are under circadian control. 

\subsection{Analysis of cluster 4}

\begin{figure}[H]
\centering\includegraphics[width=10cm]{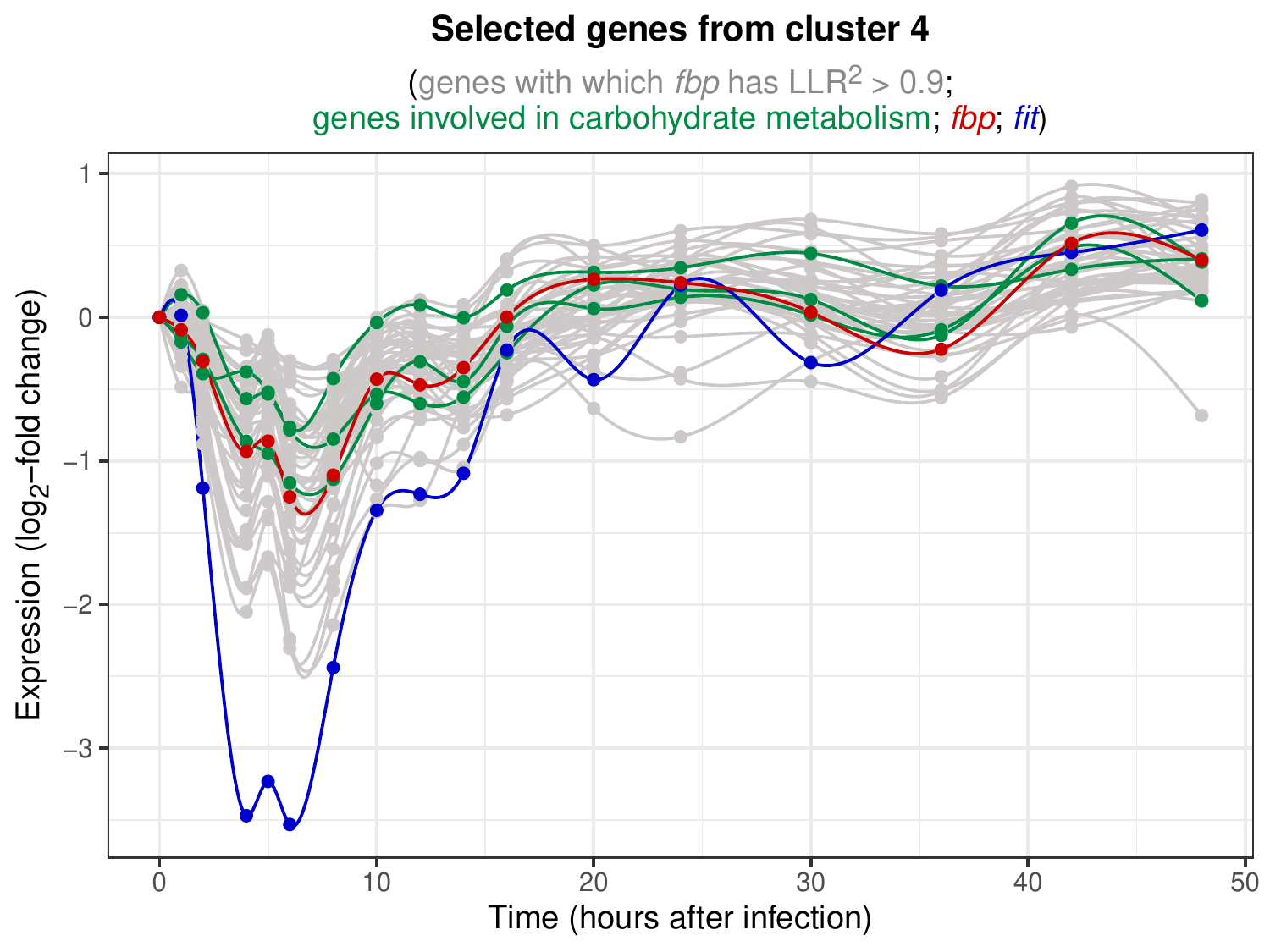}
%\end{figure}
%\begin{figure}[H]
\caption{\textit{Temporal expression patterns of selected genes in cluster 4. Light gray lines in the background correspond to genes with which the gene \textit{fbp} has a Bayesian $\LL R^2>0.9$. \textit{fbp}, shown in red, is known to be involved in carbohydrate metabolism. The three green lines correspond to genes \textit{Gale, AGBE,} and \textit{Gba1b}, which also have known roles in carbohydrate metabolism but have an unknown relationship to \textit{fbp} according to the STRING database. The dark blue line corresponds to the gene \textit{fit}, whose expression pattern is similar to that of \textit{fbp} but with much more pronounced down-regulation. \textit{fit} is known to encode a protein that stimulates insulin signaling, a process that regulates the expression of genes involved in carbohydrate metabolism.}}
\label{fig:cluster4}	
\end{figure}

\begin{figure}[H]
\centering\includegraphics[width=9.5cm]{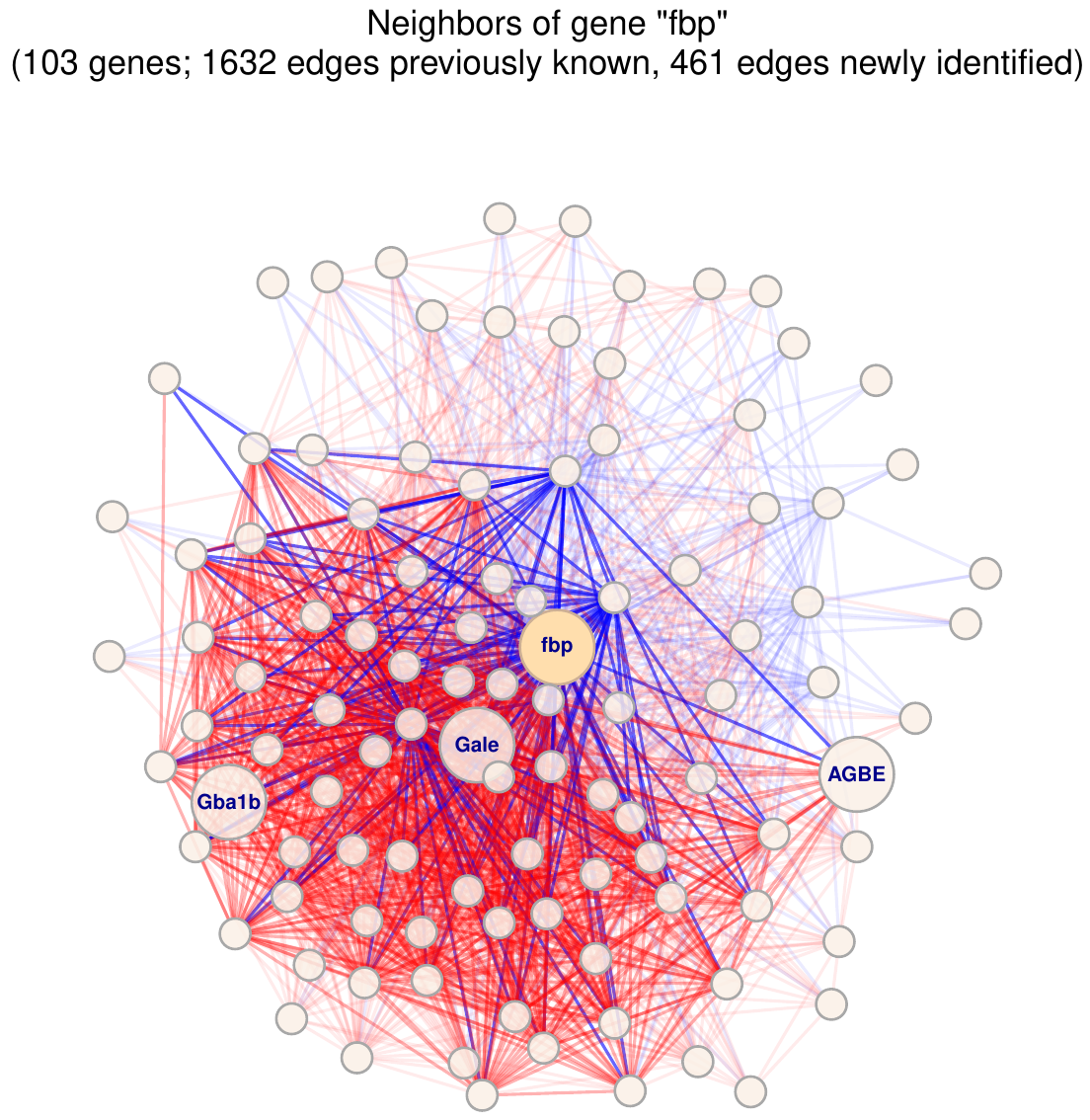}
\caption{\textit{Network of genes formed from the gene \textit{fbp} and its neighbors. Two genes are connected by an edge if their Bayesian $\LL R^2 >0.9$. Red and blue edges connect genes with known and unknown relationships, respectively. Darkened edges connect genes within cluster 4. Selected genes with uncharacterized relationships to \textit{fbp} are highlighted: \textit{Gale, AGBE, Gba1b}.}}
\label{fig:cluster4Network}	
\end{figure}

Genes in cluster 4, some of which are displayed in Figure \ref{fig:cluster4}, are characterized by a transient decrease in expression during the first 24 hours after peptidoglycan injection. This cluster was significantly enriched for ``carbohydrate metabolic process'' (B-H corrected $p$-value of $2\times 10^{-23}$). A highly-connected gene involved in carbohydrate metabolism is \textit{fbp}, which encodes the enzyme fructose-1,6-biphosphatase and has a degree of 102 in our reconstructed network shown in Figure \ref{fig:cluster4Network}. 

All of the genes to which \textit{fbp} is connected had NA values in our prior adjacency matrix, meaning that their relationship to \textit{fbp} is unknown according to our chosen sources of information. Some of these genes include \textit{Gale, AGBE,} and \textit{Gba1b}, which have known roles in carbohydrate metabolism. Twenty-two other genes connected to \textit{fbp} in our network are not well-studied. These connections suggest roles for these uncharacterized genes in carbohydrate metabolism or energy homeostasis. Another gene connected to \textit{fbp} is \textit{fit}, which has a similar expression profile as \textit{fbp} but experiences a much stronger and sharper down- and up-regulation. \textit{fit} is not directly involved in carbohydrate processing but encodes a secreted protein that stimulates insulin signaling, which in turn regulates the expression of genes involved in carbohydrate metabolism, such as \textit{fbp} \citep{sun2017drosophila}. A previous study also showed that an immune response reduces insulin signaling in \textit{Drosphila} \citep{diangelo2009immune}.

\subsection{Analysis of cluster 6}

Cluster 6 is significantly enriched for GO terms related to metabolic processes, particularly the terms ``cellular lipid catabolic process'' and ``carbohydrate metabolic process'' (B-H corrected $p$-values of $2^{-10}$ for both). In addition to these metabolic GO terms, there is significant enrichment of genes involved in ``phagocytosis'' (B-H corrected $p$-value of 0.03). During an immune response, phagocytosis is the process by which an immune cell engulfs and digests bacteria and apoptotic cells as a way to fight the infection. In \textit{Drosophila}, phagocytosis is carried out by specialized hemocytes (\textit{Drosophila} blood cells). 

Among the genes in cluster 6 involved in carbohydrate metabolism are five mannosidases (\textit{LManI, LManIII, LManIV, LManV, LManVI}) and three maltases (\textit{Mal-A2, Mal-A3, Mal-A4}). These genes encode enzymes that break down complex sugars into simple sugars like glucose. Six genes in cluster 6 are expressed in hemocytes and are involved in phagocytosis. These include four genes that belong to the Nimrod gene family (\textit{NimB4, NimC1, NimC2, eater}); \textit{Hml}, which is involved in the clotting reaction in larvae \citep{goto2003drosophila}; and \textit{Gs1}. \textit{Gs1} encodes a glutamine synthetase, an enzyme whose action is not unique to hemocytes, but that has been shown to support hemocyte function \citep{gonzalez2013glutamate}.

\begin{figure}[H]
\centering\includegraphics[width=10.5cm]{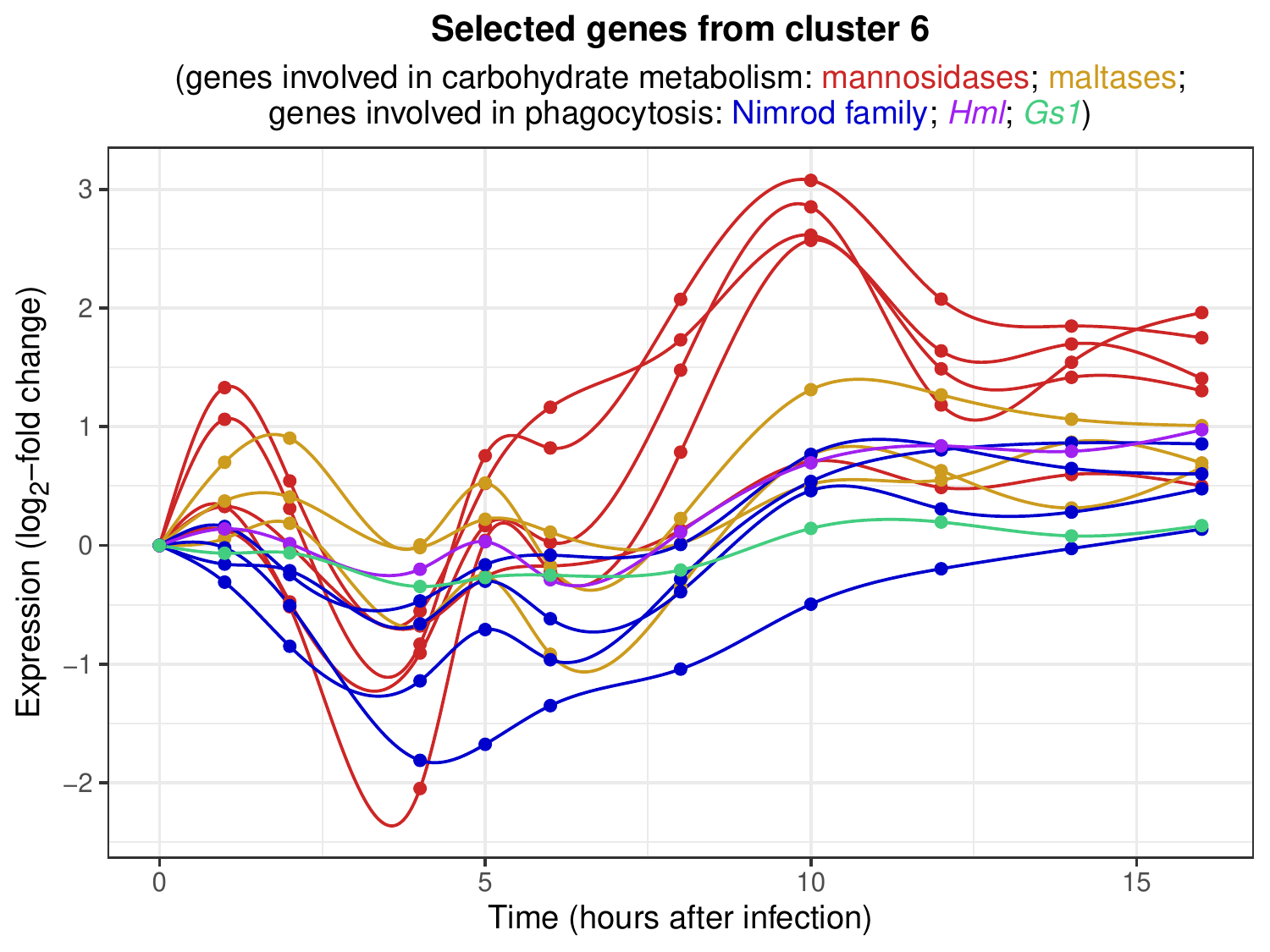}
\caption{\textit{Temporal expression patterns of selected genes in cluster 6 during the first 16 hours after peptidoglycan injection. The five red lines and three yellow lines correspond to genes involved in carbohydrate metabolism: respectively, mannosidases (\textit{LMan1, LManIII, LManIV, LManV, LManVI}) and maltases (\textit{Mal-A2, Mal-A3, Mal-A4}). The remaining lines correspond to genes that are expressed in hemocytes and are involved in phagocytosis: in blue are genes belonging to the Nimrod family (\textit{NimB4, NimC1, NimC3, eater}), in purple is \textit{Hml}, and in green is \textit{Gs1}.}}
\label{fig:cluster6}	
\end{figure}

Figure \ref{fig:cluster6} shows that the genes involved in metabolic processes and in phagocytosis exhibit similar expression patterns, with coordinated up- and down-regulation. This coordinated expression is sensible in the context of known hemocyte biology. After an infection, hemocytes undergo a metabolic switch, whereby their energy production is sustained mostly by aerobic glycolysis rather than oxidative phosphorylation \citep{krejvcova2019drosophila}. Since aerobic glycolysis is dependent on glucose, the simultaneous up-regulation of glucose-producing enzymes and genes needed for phagocytosis is aligned with our expectations. It is also worth noting that the fold changes of genes involved in phagocytosis are generally small, e.g. less than two-fold up-regulation, which is often used as a minimal cutoff in RNA-seq analyses. However, the coordinated expression changes detected by the Bayesian $\LL R^2$ suggest that these are biologically relevant patterns.    

% --- Bibliography ---

\bibliography{TeX/References}

\begin{thebibliography}{60}
\providecommand{\natexlab}[1]{#1}
\providecommand{\url}[1]{\texttt{#1}}
\expandafter\ifx\csname urlstyle\endcsname\relax
  \providecommand{\doi}[1]{doi: #1}\else
  \providecommand{\doi}{doi: \begingroup \urlstyle{rm}\Url}\fi

\bibitem[Ashburner et~al.(2000)Ashburner, Ball, Blake, Botstein, Butler,
  Cherry, Davis, Dolinski, Dwight, Eppig, et~al.]{ashburner2000gene}
M.~Ashburner, C.~A. Ball, J.~A. Blake, D.~Botstein, H.~Butler, J.~M. Cherry,
  A.~P. Davis, K.~Dolinski, S.~S. Dwight, J.~T. Eppig, et~al.
\newblock Gene {O}ntology: tool for the unification of biology.
\newblock \emph{Nature Genetics}, 25\penalty0 (1):\penalty0 25--29, 2000.

\bibitem[Bansal et~al.(2006)Bansal, Gatta, and di~Bernardo]{bansal2006}
M.~Bansal, G.~D. Gatta, and D.~di~Bernardo.
\newblock Inference of gene regulatory networks and compound mode of action
  from time course gene expression profiles.
\newblock \emph{Bioinformatics}, 22\penalty0 (7):\penalty0 815--822, 01 2006.
\newblock ISSN 1367-4803.
\newblock \doi{10.1093/bioinformatics/btl003}.

\bibitem[Bar-Joseph et~al.(2012)Bar-Joseph, Gitter, and Simon]{bar2012studying}
Z.~Bar-Joseph, A.~Gitter, and I.~Simon.
\newblock Studying and modelling dynamic biological processes using time-series
  gene expression data.
\newblock \emph{Nature Reviews Genetics}, 13\penalty0 (8):\penalty0 552--564,
  2012.

\bibitem[Benjamini and Hochberg(1995)]{benjamini1995controlling}
Y.~Benjamini and Y.~Hochberg.
\newblock Controlling the false discovery rate: a practical and powerful
  approach to multiple testing.
\newblock \emph{Journal of the Royal Statistical Society: Series B
  (Methodological)}, 57\penalty0 (1):\penalty0 289--300, 1995.

\bibitem[Carbon et~al.(2021)Carbon, Douglass, Good, Unni, Harris, Mungall,
  Basu, Chisholm, Dodson, Hartline, et~al.]{carbon2021gene}
S.~Carbon, E.~Douglass, B.~M. Good, D.~R. Unni, N.~L. Harris, C.~J. Mungall,
  S.~Basu, R.~L. Chisholm, R.~J. Dodson, E.~Hartline, et~al.
\newblock The {G}ene {O}ntology resource: enriching a {GO}ld mine.
\newblock \emph{Nucleic Acids Research}, 49\penalty0 (D1):\penalty0 D325--D334,
  2021.

\bibitem[Chen et~al.(1999)Chen, He, and Church]{chen1999modeling}
T.~Chen, H.~L. He, and G.~M. Church.
\newblock Modeling gene expression with differential equations.
\newblock In \emph{Biocomputing'99}, pages 29--40. World Scientific, 1999.

\bibitem[Clemmons et~al.(2015)Clemmons, Lindsay, and
  Wasserman]{clemmons2015effector}
A.~W. Clemmons, S.~A. Lindsay, and S.~A. Wasserman.
\newblock An effector peptide family required for {D}rosophila {T}oll-mediated
  immunity.
\newblock \emph{PLoS Pathogens}, 11\penalty0 (4):\penalty0 e1004876, 2015.

\bibitem[Cui and George(2008)]{cui2008}
W.~Cui and E.~I. George.
\newblock {Empirical Bayes vs. fully Bayes variable selection}.
\newblock \emph{Journal of Statistical Planning and Inference}, 138\penalty0
  (4):\penalty0 888--900, 2008.

\bibitem[Cyran et~al.(2003)Cyran, Buchsbaum, Reddy, Lin, Glossop, Hardin,
  Young, Storti, and Blau]{cyran2003vrille}
S.~A. Cyran, A.~M. Buchsbaum, K.~L. Reddy, M.-C. Lin, N.~R. Glossop, P.~E.
  Hardin, M.~W. Young, R.~V. Storti, and J.~Blau.
\newblock {vrille, Pdp1, and dClock form a second feedback loop in the
  Drosophila circadian clock}.
\newblock \emph{Cell}, 112\penalty0 (3):\penalty0 329--341, 2003.

\bibitem[De~Jong(2002)]{de2002modeling}
H.~De~Jong.
\newblock Modeling and simulation of genetic regulatory systems: a literature
  review.
\newblock \emph{Journal of Computational Biology}, 9\penalty0 (1):\penalty0
  67--103, 2002.

\bibitem[D'haeseleer et~al.(1999)D'haeseleer, Wen, Fuhrman, and
  Somogyi]{d1999linear}
P.~D'haeseleer, X.~Wen, S.~Fuhrman, and R.~Somogyi.
\newblock {Linear modeling of mRNA expression levels during CNS development and
  injury}.
\newblock In \emph{Biocomputing'99}, pages 41--52. World Scientific, 1999.

\bibitem[DiAngelo et~al.(2009)DiAngelo, Bland, Bambina, Cherry, and
  Birnbaum]{diangelo2009immune}
J.~R. DiAngelo, M.~L. Bland, S.~Bambina, S.~Cherry, and M.~J. Birnbaum.
\newblock The immune response attenuates growth and nutrient storage in
  {D}rosophila by reducing insulin signaling.
\newblock \emph{Proceedings of the National Academy of Sciences}, 106\penalty0
  (49):\penalty0 20853--20858, 2009.

\bibitem[Efron(2004)]{efron2004}
B.~Efron.
\newblock The estimation of prediction error: Covariance penalties and
  cross-validation.
\newblock \emph{Journal of the American Statistical Association}, 99\penalty0
  (467):\penalty0 619--632, 2004.

\bibitem[Eisen et~al.(1998)Eisen, Spellman, Brown, and
  Botstein]{eisen1998cluster}
M.~B. Eisen, P.~T. Spellman, P.~O. Brown, and D.~Botstein.
\newblock Cluster analysis and display of genome-wide expression patterns.
\newblock \emph{Proceedings of the National Academy of Sciences}, 95\penalty0
  (25):\penalty0 14863--14868, 1998.

\bibitem[Fabregat et~al.(2018)Fabregat, Jupe, Matthews, Sidiropoulos,
  Gillespie, Garapati, Haw, Jassal, Korninger, May,
  et~al.]{fabregat2018reactome}
A.~Fabregat, S.~Jupe, L.~Matthews, K.~Sidiropoulos, M.~Gillespie, P.~Garapati,
  R.~Haw, B.~Jassal, F.~Korninger, B.~May, et~al.
\newblock {The Reactome pathway knowledgebase}.
\newblock \emph{Nucleic Acids Research}, 46\penalty0 (D1):\penalty0 D649--D655,
  2018.

\bibitem[Farina et~al.(2007)Farina, {De Santis}, Morelli, and
  Ruberti]{Farina2007}
L.~Farina, A.~{De Santis}, G.~Morelli, and I.~Ruberti.
\newblock {Dynamic measure of gene co-regulation}.
\newblock \emph{IET Systems Biology}, 1\penalty0 (1):\penalty0 10--17, 2007.
\newblock ISSN 17518849.
\newblock \doi{10.1049/iet-syb:20060031}.

\bibitem[Farina et~al.(2008)Farina, {De Santis}, Salvucci, Morelli, and
  Ruberti]{Farina2008}
L.~Farina, A.~{De Santis}, S.~Salvucci, G.~Morelli, and I.~Ruberti.
\newblock {Embedding mRNA stability in correlation analysis of time-series gene
  expression data}.
\newblock \emph{PLoS Computational Biology}, 4\penalty0 (8), 2008.
\newblock ISSN 1553734X.
\newblock \doi{10.1371/journal.pcbi.1000141}.

\bibitem[Fourdrinier et~al.(2018)Fourdrinier, Strawderman, and
  Wells]{Wells2018}
D.~Fourdrinier, W.~E. Strawderman, and M.~T. Wells.
\newblock \emph{{Shrinkage Estimation}}.
\newblock Springer, 2018.

\bibitem[Gaudet et~al.(2011)Gaudet, Livstone, Lewis, and
  Thomas]{gaudet2011phylogenetic}
P.~Gaudet, M.~S. Livstone, S.~E. Lewis, and P.~D. Thomas.
\newblock Phylogenetic-based propagation of functional annotations within the
  {G}ene {O}ntology consortium.
\newblock \emph{Briefings in bioinformatics}, 12\penalty0 (5):\penalty0
  449--462, 2011.

\bibitem[Gelman et~al.(2018)Gelman, Goodrich, Gabry, and Vethari]{Gelman2018}
A.~Gelman, B.~Goodrich, J.~Gabry, and A.~Vethari.
\newblock {R-squared for Bayesian regression models}.
\newblock \emph{American Statistician}, 2018.

\bibitem[George and Foster(2000)]{GeorgeFoster2000}
E.~I. George and D.~P. Foster.
\newblock {Calibration and empirical Bayes variable selection}.
\newblock \emph{Biometrika}, 87\penalty0 (4):\penalty0 731--747, 2000.
\newblock ISSN 00063444.
\newblock \doi{10.1093/biomet/87.4.731}.

\bibitem[Giles and Rayner(1979)]{giles1979mean}
D.~Giles and A.~Rayner.
\newblock {The mean squared errors of the maximum likelihood and
  natural-conjugate Bayes regression estimators}.
\newblock \emph{Journal of Econometrics}, 11\penalty0 (2-3):\penalty0 319--334,
  1979.

\bibitem[Gobert et~al.(2003)Gobert, Gottar, Matskevich, Rutschmann, Royet,
  Belvin, Hoffmann, and Ferrandon]{gobert2003dual}
V.~Gobert, M.~Gottar, A.~A. Matskevich, S.~Rutschmann, J.~Royet, M.~Belvin,
  J.~A. Hoffmann, and D.~Ferrandon.
\newblock Dual activation of the {D}rosophila toll pathway by two pattern
  recognition receptors.
\newblock \emph{Science}, 302\penalty0 (5653):\penalty0 2126--2130, 2003.

\bibitem[Gonzalez et~al.(2013)Gonzalez, Garg, Tang, Nazario-Toole, and
  Wu]{gonzalez2013glutamate}
E.~A. Gonzalez, A.~Garg, J.~Tang, A.~E. Nazario-Toole, and L.~P. Wu.
\newblock A glutamate-dependent redox system in blood cells is integral for
  phagocytosis in {D}rosophila melanogaster.
\newblock \emph{Current Biology}, 23\penalty0 (22):\penalty0 2319--2324, 2013.

\bibitem[Goto et~al.(2003)Goto, Kadowaki, and Kitagawa]{goto2003drosophila}
A.~Goto, T.~Kadowaki, and Y.~Kitagawa.
\newblock Drosophila hemolectin gene is expressed in embryonic and larval
  hemocytes and its knock down causes bleeding defects.
\newblock \emph{Developmental Biology}, 264\penalty0 (2):\penalty0 582--591,
  2003.

\bibitem[Hanson and Lemaitre(2020)]{hanson2020new}
M.~A. Hanson and B.~Lemaitre.
\newblock New insights on {D}rosophila antimicrobial peptide function in host
  defense and beyond.
\newblock \emph{Current Opinion in Immunology}, 62:\penalty0 22--30, 2020.

\bibitem[Hill et~al.(2012)Hill, Neve, Bayani, Kuo, Ziyad, Spellman, Gray, and
  Mukherjee]{Hill2012}
S.~M. Hill, R.~M. Neve, N.~Bayani, W.~L. Kuo, S.~Ziyad, P.~T. Spellman, J.~W.
  Gray, and S.~Mukherjee.
\newblock {Integrating biological knowledge into variable selection: an
  empirical Bayes approach with an application in cancer biology}.
\newblock \emph{BMC Bioinformatics}, 13\penalty0 (1):\penalty0 1--15, 2012.
\newblock ISSN 14712105.
\newblock \doi{10.1186/1471-2105-13-94}.

\bibitem[Kanehisa and Goto(2000)]{kanehisa2000kegg}
M.~Kanehisa and S.~Goto.
\newblock {KEGG: Kyoto encyclopedia of genes and genomes}.
\newblock \emph{Nucleic Acids Research}, 28\penalty0 (1):\penalty0 27--30,
  2000.

\bibitem[Kaneko et~al.(2004)Kaneko, Goldman, Mellroth, Steiner, Fukase,
  Kusumoto, Harley, Fox, Golenbock, and Silverman]{kaneko2004monomeric}
T.~Kaneko, W.~E. Goldman, P.~Mellroth, H.~Steiner, K.~Fukase, S.~Kusumoto,
  W.~Harley, A.~Fox, D.~Golenbock, and N.~Silverman.
\newblock Monomeric and polymeric gram-negative peptidoglycan but not purified
  {LPS} stimulate the {D}rosophila {IMD} pathway.
\newblock \emph{Immunity}, 20\penalty0 (5):\penalty0 637--649, 2004.

\bibitem[Karp et~al.(2019)Karp, Billington, Caspi, Fulcher, Latendresse,
  Kothari, Keseler, Krummenacker, Midford, Ong, et~al.]{karp2019biocyc}
P.~D. Karp, R.~Billington, R.~Caspi, C.~A. Fulcher, M.~Latendresse, A.~Kothari,
  I.~M. Keseler, M.~Krummenacker, P.~E. Midford, Q.~Ong, et~al.
\newblock {The BioCyc collection of microbial genomes and metabolic pathways}.
\newblock \emph{Briefings in Bioinformatics}, 20\penalty0 (4):\penalty0
  1085--1093, 2019.

\bibitem[Kass and Raftery(1995)]{kass1995bayes}
R.~E. Kass and A.~E. Raftery.
\newblock Bayes factors.
\newblock \emph{Journal of the American Statistical Association}, 90\penalty0
  (430):\penalty0 773--795, 1995.

\bibitem[Krej{\v{c}}ov{\'a} et~al.(2019)Krej{\v{c}}ov{\'a}, Danielov{\'a},
  Nedbalov{\'a}, Kazek, Strych, Chawla, Tennessen, Lieskovsk{\'a}, Jindra,
  Dole{\v{z}}al, et~al.]{krejvcova2019drosophila}
G.~Krej{\v{c}}ov{\'a}, A.~Danielov{\'a}, P.~Nedbalov{\'a}, M.~Kazek, L.~Strych,
  G.~Chawla, J.~M. Tennessen, J.~Lieskovsk{\'a}, M.~Jindra, T.~Dole{\v{z}}al,
  et~al.
\newblock Drosophila macrophages switch to aerobic glycolysis to mount
  effective antibacterial defense.
\newblock \emph{Elife}, 8:\penalty0 e50414, 2019.

\bibitem[Larkin et~al.(2020)Larkin, Marygold, Antonazzo, Attrill, Dos~Santos,
  Garapati, Goodman, Gramates, Millburn, Strelets, et~al.]{larkin2020flybase}
A.~Larkin, S.~J. Marygold, G.~Antonazzo, H.~Attrill, G.~Dos~Santos, P.~V.
  Garapati, J.~L. Goodman, L.~S. Gramates, G.~Millburn, V.~B. Strelets, et~al.
\newblock {FlyBase: updates to the Drosophila melanogaster knowledge base}.
\newblock \emph{Nucleic Acids Research}, 2020.

\bibitem[Law et~al.(2014)Law, Chen, Shi, and Smyth]{law2014voom}
C.~W. Law, Y.~Chen, W.~Shi, and G.~K. Smyth.
\newblock {voom: Precision weights unlock linear model analysis tools for
  RNA-seq read counts}.
\newblock \emph{Genome Biology}, 15\penalty0 (2):\penalty0 R29, 2014.

\bibitem[Li and Zhang(2010)]{li2010bayesian}
F.~Li and N.~R. Zhang.
\newblock Bayesian variable selection in structured high-dimensional covariate
  spaces with applications in genomics.
\newblock \emph{Journal of the American Statistical Association}, 105\penalty0
  (491):\penalty0 1202--1214, 2010.

\bibitem[Li and Jackson(2015)]{li2015gene}
Y.~Li and S.~A. Jackson.
\newblock Gene network reconstruction by integration of prior biological
  knowledge.
\newblock \emph{G3: Genes, Genomes, Genetics}, 5\penalty0 (6):\penalty0
  1075--1079, 2015.

\bibitem[Liang et~al.(2008)Liang, Paulo, Molina, Clyde, and Berger]{Liang2008}
F.~Liang, R.~Paulo, G.~Molina, M.~A. Clyde, and J.~O. Berger.
\newblock {Mixtures of g priors for Bayesian variable selection}.
\newblock \emph{Journal of the American Statistical Association}, 103\penalty0
  (481):\penalty0 410--423, 2008.
\newblock ISSN 01621459.
\newblock \doi{10.1198/016214507000001337}.

\bibitem[Lo et~al.(2012)Lo, Raftery, Dombek, Zhu, Schadt, Bumgarner, and
  Yeung]{Lo2012}
K.~Lo, A.~E. Raftery, K.~M. Dombek, J.~Zhu, E.~E. Schadt, R.~E. Bumgarner, and
  K.~Y. Yeung.
\newblock {Integrating external biological knowledge in the construction of
  regulatory networks from time-series expression data}.
\newblock \emph{BMC Systems Biology}, 6, 2012.
\newblock ISSN 17520509.
\newblock \doi{10.1186/1752-0509-6-101}.

\bibitem[Nepomuceno et~al.(2015)Nepomuceno, Troncoso, Nepomuceno-Chamorro, and
  Aguilar-Ruiz]{Nepomuceno2015}
J.~A. Nepomuceno, A.~Troncoso, I.~A. Nepomuceno-Chamorro, and J.~S.
  Aguilar-Ruiz.
\newblock {Integrating biological knowledge based on functional annotations for
  biclustering of gene expression data}.
\newblock \emph{Computer Methods and Programs in Biomedicine}, 119\penalty0
  (3):\penalty0 163--180, 2015.
\newblock ISSN 18727565.
\newblock \doi{10.1016/j.cmpb.2015.02.010}.

\bibitem[Peng et~al.(2013)Peng, Zhu, Ander, Zhang, Xue, Sharp, and
  Yang]{peng2013integrative}
B.~Peng, D.~Zhu, B.~P. Ander, X.~Zhang, F.~Xue, F.~R. Sharp, and X.~Yang.
\newblock {An integrative framework for Bayesian variable selection with
  informative priors for identifying genes and pathways}.
\newblock \emph{PloS One}, 8\penalty0 (7):\penalty0 e67672, 2013.

\bibitem[Polynikis et~al.(2009)Polynikis, Hogan, and
  di~Bernardo]{polynikis2009comparing}
A.~Polynikis, S.~Hogan, and M.~di~Bernardo.
\newblock {Comparing different ODE modelling approaches for gene regulatory
  networks}.
\newblock \emph{Journal of Theoretical Biology}, 261\penalty0 (4):\penalty0
  511--530, 2009.

\bibitem[Purvis and Lahav(2013)]{purvis2013encoding}
J.~E. Purvis and G.~Lahav.
\newblock Encoding and decoding cellular information through signaling
  dynamics.
\newblock \emph{Cell}, 152\penalty0 (5):\penalty0 945--956, 2013.

\bibitem[Qian et~al.(2001)Qian, Dolled-Filhart, Lin, Yu, and
  Gerstein]{qian2001beyond}
J.~Qian, M.~Dolled-Filhart, J.~Lin, H.~Yu, and M.~Gerstein.
\newblock Beyond synexpression relationships: local clustering of time-shifted
  and inverted gene expression profiles identifies new, biologically relevant
  interactions.
\newblock \emph{Journal of Molecular Biology}, 314\penalty0 (5):\penalty0
  1053--1066, 2001.

\bibitem[Research()]{Mathematica}
W.~Research.
\newblock Mathematica, {V}ersion 12.0.
\newblock Champaign, IL, 2019.

\bibitem[Schlamp et~al.(2021)Schlamp, Delbare, Early, Wells, Basu, and
  Clark]{schlamp2020}
F.~Schlamp, S.~Y. Delbare, A.~M. Early, M.~T. Wells, S.~Basu, and A.~G. Clark.
\newblock {Dense time-course gene expression profiling of the Drosophila
  melanogaster innate immune response}.
\newblock \emph{BMC genomics}, 22\penalty0 (1):\penalty0 1--22, 2021.

\bibitem[Stingo et~al.(2011)Stingo, Chen, Tadesse, and Vannucci]{stingo2011}
F.~C. Stingo, Y.~A. Chen, M.~G. Tadesse, and M.~Vannucci.
\newblock Incorporating biological information into linear models: A {B}ayesian
  approach to the selection of pathways and genes.
\newblock \emph{The Annals of Applied Statistics}, 5\penalty0 (3), 2011.

\bibitem[Sun et~al.(2017)Sun, Liu, Bai, Li, Li, Zhang, Zhang, Guo, and
  Li]{sun2017drosophila}
J.~Sun, C.~Liu, X.~Bai, X.~Li, J.~Li, Z.~Zhang, Y.~Zhang, J.~Guo, and Y.~Li.
\newblock Drosophila {FIT} is a protein-specific satiety hormone essential for
  feeding control.
\newblock \emph{Nature communications}, 8\penalty0 (1):\penalty0 1--13, 2017.

\bibitem[Szklarczyk et~al.(2019)Szklarczyk, Gable, Lyon, Junge, Wyder,
  Huerta-Cepas, Simonovic, Doncheva, Morris, Bork,
  et~al.]{szklarczyk2019string}
D.~Szklarczyk, A.~L. Gable, D.~Lyon, A.~Junge, S.~Wyder, J.~Huerta-Cepas,
  M.~Simonovic, N.~T. Doncheva, J.~H. Morris, P.~Bork, et~al.
\newblock {STRING v11: protein--protein association networks with increased
  coverage, supporting functional discovery in genome-wide experimental
  datasets}.
\newblock \emph{Nucleic Acids Research}, 47\penalty0 (D1):\penalty0 D607--D613,
  2019.

\bibitem[Tavazoie et~al.(1999)Tavazoie, Hughes, Campbell, Cho, and
  Church]{tavazoie1999systematic}
S.~Tavazoie, J.~D. Hughes, M.~J. Campbell, R.~J. Cho, and G.~M. Church.
\newblock Systematic determination of genetic network architecture.
\newblock \emph{Nature Genetics}, 22\penalty0 (3):\penalty0 281--285, 1999.

\bibitem[Valanne et~al.(2019)Valanne, Salminen, J{\"a}rvel{\"a}-St{\"o}lting,
  Vesala, and R{\"a}met]{valanne2019immune}
S.~Valanne, T.~S. Salminen, M.~J{\"a}rvel{\"a}-St{\"o}lting, L.~Vesala, and
  M.~R{\"a}met.
\newblock Immune-inducible non-coding {RNA} molecule linc{RNA-IBIN} connects
  immunity and metabolism in drosophila melanogaster.
\newblock \emph{PLoS Pathogens}, 15\penalty0 (1):\penalty0 e1007504, 2019.

\bibitem[Wang et~al.(2010)Wang, Wang, Jiao, von Lintig, and
  Montell]{wang2010requirement}
X.~Wang, T.~Wang, Y.~Jiao, J.~von Lintig, and C.~Montell.
\newblock Requirement for an enzymatic visual cycle in {D}rosophila.
\newblock \emph{Current Biology}, 20\penalty0 (2):\penalty0 93--102, 2010.

\bibitem[Ward~Jr(1963)]{ward1963hierarchical}
J.~H. Ward~Jr.
\newblock Hierarchical grouping to optimize an objective function.
\newblock \emph{Journal of the American Statistical Association}, 58\penalty0
  (301):\penalty0 236--244, 1963.

\bibitem[Wu et~al.(2019)Wu, Qiu, xiang Yuan, and Wu]{wu2019ODE}
L.~Wu, X.~Qiu, Y.~xiang Yuan, and H.~Wu.
\newblock Parameter estimation and variable selection for big systems of linear
  ordinary differential equations: A matrix-based approach.
\newblock \emph{Journal of the American Statistical Association}, 114\penalty0
  (526):\penalty0 657--667, 2019.
\newblock \doi{10.1080/01621459.2017.1423074}.

\bibitem[Yosef and Regev(2011)]{yosef2011impulse}
N.~Yosef and A.~Regev.
\newblock {Impulse control: temporal dynamics in gene transcription}.
\newblock \emph{Cell}, 144\penalty0 (6):\penalty0 886--896, 2011.

\bibitem[Yu et~al.(2010)Yu, Li, Qin, Bo, Wu, and Wang]{yu2010gosemsim}
G.~Yu, F.~Li, Y.~Qin, X.~Bo, Y.~Wu, and S.~Wang.
\newblock Go{S}em{S}im: an {R} package for measuring semantic similarity among
  go terms and gene products.
\newblock \emph{Bioinformatics}, 26\penalty0 (7):\penalty0 976--978, 2010.

\bibitem[Yu et~al.(2012)Yu, Wang, Han, and He]{yu2012clusterprofiler}
G.~Yu, L.-G. Wang, Y.~Han, and Q.-Y. He.
\newblock cluster{P}rofiler: an {R} package for comparing biological themes
  among gene clusters.
\newblock \emph{Omics: A Journal of Integrative Biology}, 16\penalty0
  (5):\penalty0 284--287, 2012.

\bibitem[Zaidman-R{\'e}my et~al.(2011)Zaidman-R{\'e}my, Poidevin, Herv{\'e},
  Welchman, Paredes, Fahlander, Steiner, Mengin-Lecreulx, and
  Lemaitre]{zaidman2011drosophila}
A.~Zaidman-R{\'e}my, M.~Poidevin, M.~Herv{\'e}, D.~P. Welchman, J.~C. Paredes,
  C.~Fahlander, H.~Steiner, D.~Mengin-Lecreulx, and B.~Lemaitre.
\newblock Drosophila immunity: analysis of {PGRP-SB1} expression, enzymatic
  activity and function.
\newblock \emph{PLoS One}, 6\penalty0 (2):\penalty0 e17231, 2011.

\bibitem[Zellner(1986)]{zellner1986}
A.~Zellner.
\newblock {On assessing prior distributions and Bayesian regression analysis
  with g-prior distributions}.
\newblock \emph{Bayesian Inference and Decision Techniques}, 1986.

\bibitem[Zellner and Siow(1980)]{Zellner1980}
A.~Zellner and A.~Siow.
\newblock {Posterior odds ratios for selected regression hypotheses}.
\newblock \emph{Trabajos de Estadistica Y de Investigacion Operativa},
  31\penalty0 (1):\penalty0 585--603, 1980.
\newblock ISSN 00410241.
\newblock \doi{10.1007/BF02888369}.

\bibitem[Zhang et~al.(2013)Zhang, Wan, Allen, Pang, Anderson, and
  Liu]{zhang2013molecular}
W.~Zhang, Y.-w. Wan, G.~I. Allen, K.~Pang, M.~L. Anderson, and Z.~Liu.
\newblock Molecular pathway identification using biological network-regularized
  logistic models.
\newblock \emph{BMC Genomics}, 14\penalty0 (8):\penalty0 1--8, 2013.

\end{thebibliography}
\bibliographystyle{abbrvnat}

% --- Document ends here ---

\end{document}